\documentstyle[12pt]{article}
\def\today{\number\day\space
     \ifcase\month\or
       January\or February\or March\or April\or May\or June\or
       July\or August\or September\or October\or November\or December\fi
     \space\number\year}
\newcommand{\be}{\begin{equation}}
\newcommand{\ee}{\end{equation}}
\newcommand{\ba}{\begin{eqnarray}}
\newcommand{\ea}{\end{eqnarray}}
\newcommand{\dem}[1]{\Delta \left. m_{#1}^2\right|_{em}}
\newcommand{\tr}{{\rm tr}}
\begin{document}
\begin{titlepage}
\begin{flushright}
NORDITA 93/15 N,P\\
revised version
\end{flushright}
\vfill
\begin{center}
{\large\bf Violations of Dashen's Theorem}\\[3cm]
Johan Bijnens\\[0.5cm]
NORDITA, Blegdamsvej 17\\
DK-2100 Copenhagen \O, Denmark
\end{center}
\vfill
\begin{abstract}
The electromagnetic contribution to the $K^+ - K^0$ mass difference is
calculated in the $1/N_c$ approach including the $SU(3)$ breaking contributions
due to a non-zero strange quark mass. The short-distance contribution
can be unambiguously determined in terms of the known parameters of the
next-to-leading order chiral lagrangian. The long distance part is estimated
using the $1/N_c$ approach by Bardeen et al.
Relatively large corrections are found.
\end{abstract}
\vfill
\end{titlepage}

{\bf 1) Introduction:}
The values of the quark masses are important physical quantities. Determining
the values of the quark masses is, however, not such a simple task because
quarks are confined and do not exist as free particles whose mass can be
measured directly. At high energies they can be identified with jets but
the effects due to their masses at these high energies are so small that
jet-physics cannot be used to determine the masses of the light quarks.
Therefore a recourse to other methods is needed. For a review of the
values of the quark masses see \cite{GLPR}. A more recent discussion of some
of the issues arising in this context can be found in \cite{L}.

Good theoretical results exist for certain ratios of the quark masses
in terms of the masses of the pseudoscalar meson octet, $\pi$, $K$ and $\eta$.
As an example we give here the relation that determines the isospin
breaking parameter $m_u - m_d$, valid to next-to-leading order \cite{GL}.
\ba
\label{msmd}
\frac{m_d - m_u}{m_s - \hat{m}}\frac{2\hat{m}}{m_s + \hat{m}}
= \frac{m_{K^0}^2 - m_{K^+}^2 - m_{\pi^0}^2 + m_{\pi^+}^2}{m_K^2 - M_\pi^2}
\frac{m_\pi^2}{m_K^2}\times
\nonumber\\
\left\{1 + O(m_q^2) + O\left( e^2 \frac{m_s - \hat{m}}{m_d - m_u} \right)
\right\}\ .
\ea
The difference of meson masses in the numerator removes the leading
electromagnetic isospin breaking contribution to the masses, the leading
electromagnetic correction to the difference of the mass squared of kaons
and pions are equal\cite{dashen}. This is known as Dashen's theorem.
Corrections
to it are of the order of the last term in eq. (\ref{msmd}).

In this letter we evaluate the last type of correction within the framework
of the $1/N_c$ approach. This approach was first applied to hadronic matrix
elements needed for non-leptonic kaon decays by Bardeen et al.\cite{BBG}.
The same approach was then used to calculate the electromagnetic
correction to the pion masses in the chiral limit\cite{BBG2}. In this
case because of the presence of the photon the matching problem could
be circumvented and the general approach tested. Good agreement with
the observed pion mass difference was obtained.
This method was then extended to the calculation of the
long distance part of the $K_L -K_S$ mass difference. Again
a reasonable matching and reasonable agreement with
the measured mass difference was found\cite{BG}. This has prompted the
investigation of the term of order $e^2 m_q$
in the electromagnetic mass differences using the same approach.

In the case of the $\pi^+-\pi^0$ mass difference the smallness of the quark
masses allows the use of current algebra and the mass difference can be
related to a dispersion integral over the difference of the vector and
axial-vector two-point functions \cite{das}. This method cannot be
used in the present case since the non-zero value of the quark mass gives
corrections to the above relation. Here we directly evaluate the mass
differences from the electromagnetic effective
action. To lowest order in $e^2$
the electromagnetic contribution to the mass squared of
meson $M$ is given by
\be
\label{eq2}
\left. m_M^2\right|_{em}
 = - \langle M | e^2 \int \frac{d^4 q}{16\pi^4}
\frac{J_\mu(q)J_\nu(-q)}{q^2}\left(g^{\mu\nu}-\xi\frac{q^\mu q^\nu}{q^2}
\right)
 | M \rangle\ ,
\ee
where $e$ is the electromagnetic charge unit and $J_\mu(q)$ is
the electromagnetic current. The factor $q^{-2}$ comes from the photon
propagator
and we work in an arbitrary gauge (the final results are $\xi$-independent).
The matrix element in (\ref{eq2}) has to be evaluated in the presence of
the strong interaction only.

We now perform the analytic continuation to euclidean space and split
the integral in (\ref{eq2}) in two parts. One with $q^2 \ge \Lambda^2$
and one with $q^2 \le \Lambda^2$. We call the scale $\Lambda$ the
matching scale.
The former part has a high momentum photon.
There we use the couplings of the photons to quarks to derive an expression
in terms of local quark operators. The expectation values of these
local quark operators are then determined using the $1/N_c$ approach.
The part with $q^2 \le \Lambda^2$ has a low momentum photon and cannot be
evaluated using that method. This part we evaluate using an effective
Lagrangian description of the low-energy hadronic interactions.

For this effective Lagrangian we use the lowest dimension chiral Lagrangian.
In ref. \cite{BBG2} substantially better matching between the low and high
momentum domain was achieved by including vector and axial-vector mesons.
To do this here requires a correct description of $SU(3)$ breaking in the
vector and axial-vector sector. This does not exist at present. However,
the numerical result only changed by a rather small amount after inclusion
of these effects therefore we expect a similar result here.\footnote{
As an example the coupling of photons to the mesons will also have corrections
due to the quark mass. These are not included at present. They only appear
at $O(p^6)$ in the chiral Lagrangian at leading $1/N_c$.}
All results are quoted to first order in $m_s \ne 0$ and
we set $m_u = m_d = 0$.

{\bf 2) Short distance :} First we derive the effective action involving
quarks doing the integral over photon momenta for $q^2$ larger than
$\Lambda^2$. There are to order $\alpha_S$ three types of contributions.
The type of diagrams involved in these processes are shown in fig. \ref{fig1}.

The first type of diagrams, fig. \ref{fig1}a, only contributes via the
electromagnetic renormalization of the current quark mass. This type of
contribution is logarithmically divergent and the ultraviolet divergence
has to be removed using counterterms. If we have defined the current quark
mass at a scale $\mu_0$ this set of diagrams results in an effective
action of the type
\be
iS_{eff}^1 = \frac{-i\alpha_{em}}{12\pi} m_s \overline{s}s
\log\frac{\mu_0^2}{\Lambda^2} + O\left( \alpha_{em}\alpha_S \right) \ .
\ee
This type of contribution directly influences the kaon mass via the standard
lowest order relation
\be
\label{mkvev}
m_K^2 \approx 2 m_s |\langle \overline{s}s \rangle| / f^2 \ ,
\ee
since $S^1_{eff}$ corresponds to a shift in $m_s$ to be included in
(\ref{mkvev}).
We work in a normalization where $f\approx 132~MeV$.
Its effects contribute the same to the $K^+$ and the $K^0$ to the order we
are working so it doesn't contribute to the difference.

The second type of diagrams, which we could describe as the photonic
penguin, does not contribute in the chiral limit. It can only
produce terms of isospin at most one, since it only involves two quarks.
The mass difference between $\pi^+$ and $\pi^0$ is isospin 2 so this
type of diagrams cannot contribute to it and not to $\dem{K}$ by using
$SU(3)$. It can contribute as soon as $SU(3)$ is broken.
The effective action resulting from the diagrams of fig. \ref{fig1}b
is~:
\be
\label{penguin}
iS_{eff}^2  =  \frac{i\alpha_{em}\alpha_S}{27\Lambda^2}
\left\{  4 \overline{u}u + \overline{d}d + \overline{s}s\right\}_{V\alpha\beta}
\left\{   \overline{u}u + \overline{d}d + \overline{s}s\right\}_{V\beta\alpha}
\ee
with
$\{\overline{q}q\}_{V\alpha\beta} = \overline{q}_\alpha \gamma_\mu q_\beta$
and $\alpha$ and $\beta$ are colour indices and are summed over.
There are no terms of $O(m_s)$ in this expression. The $SU(3)$ breaking due
to the quark masses appears only at higher order in this type of diagrams.

The third type of diagrams, fig. \ref{fig1}c and permutations or the box
diagrams, are the only ones that contribute in the chiral limit. Again there
are no correction terms of $O(m_s)$ in this set of contributions.
The resulting effective action is
\be
\label{box}
iS_{eff}^3 = \frac{-i\alpha_{em}\alpha_S}{6\Lambda^2}
\left\{-4(uu)-(dd)-(ss)+4(ud)+4(us)-2(ds)\right\}\ ,
\ee
with $(qq') = (\overline{q}_\alpha\gamma_\mu\gamma_5 q_\beta)
(\overline{q'}_\beta\gamma_\mu\gamma_5 q'_\alpha )$ and again $\alpha,\beta$
are colour indices and are summed over.
We have used here large $N_c$ relations between the colour matrices and
dropped the terms which are suppressed by $1/N_c$.

To evaluate the matrix
elements between two mesons of these effective actions we now
use the Fierz relations to bring them in the general form of
products of two colour singlet combinations. In the large $N_c$ limit these
colour singlets hadronize independently and we can use standard chiral
perturbation theory in the large $N_c$ limit to evaluate them.
The relevant terms here are the two lowest order terms and the ones
proportional to $L_5,~L_8$ and $H_2$, see ref. \cite{GL}.
The Fierzed version of eqs.(\ref{penguin},\ref{box}) contain squares
of vector, axial-vector, pseudoscalar and scalar densities.
The vector densities squared
only contribute to processes with at least four
pseudoscalars and cannot contribute to the mass differences.
The axial-vector density contains at least one derivative which means that
its contribution always comes multiplied with the mass squared of the meson
involved. This mass is itself already of order $m_s$ so we only need
the leading term.
\be
\overline{q}\gamma_\mu\gamma_5 T^a q = f \tr \partial_\mu
T^a M + \cdots\ .
\ee
This is in terms of the meson matrix
\be
M = \pmatrix{\frac{\pi^0}{\sqrt{2}}+\frac{\eta}{\sqrt{6}} &\pi^+ & K^+\cr
    \pi^- & \frac{-\pi^0}{\sqrt{2}}+\frac{\eta}{\sqrt{6}} & K^0\cr
   K^- & \overline{K^0} & \frac{-2\eta}{\sqrt{6}} } \ ,
\ee
and a flavour matrix $T^a$ and $\overline{q} =
\left(\overline{u}\ \overline{d}\ \overline{s}\right)$.
The scalar and pseudoscalar densities already contribute at leading
order and are hence needed to second order.
The parameters $B_0$, $L_5$, $L_8$ and $H_2$ are those of ref.\cite{GL}.
The quark mass contribution is contained in $M_Q = {\rm diag}(0,0,m_s)$.
Only the terms up to $O(M^2)$ are given.
\ba
\overline{q}T^a q &=& \frac{-B_0 f^2}{2}\tr T^a
 - 8B_0^2 (2L_8 +H_2)\tr T^a M_Q + B_0 \tr T^a M^2
\nonumber           \\
& &-\frac{16B_0 L_5}{f^2}\tr T^a \partial_\mu M \partial^\mu M
  +32B_0^2 L_8 \tr T^a\left\{ M,\left\{ M, M_Q  \right\}\right\} \ ,
  \nonumber\\
\overline{q}\gamma_5 T^a q &=& iB_0 f \tr T^a M
+ \frac{32B_0^2 L_8}{f}\tr T^a\left(M_Q M + M M_Q\right)\ .
\ea

Putting all the above formulas together leads to the following short distance
electromagnetic contributions to masses in the limit where
$m_u = m_d =0$ and $m_s \ne 0$~:
\ba
\label{SD}
\left. m_{\pi^0}^2\right|_{em}^{SD} &=& 0\ ,\nonumber\\
\left. m_{\pi^+}^2\right|_{em}^{SD} &=&
 \frac{3\alpha_S\alpha_{em}}{2\Lambda^2}B_0^2 f^2 \ ,\nonumber\\
\left. m_{K^0}^2\right|_{em}^{SD} &=&
\frac{\alpha_{em}\alpha_S}{\Lambda^2}\Bigg\{
\left(\frac{-16}{3} +\frac{32}{27}\right) B_0^2 L_5 m_K^2 +
\left(\frac{-1}{6}-\frac{1}{27}\right) f^2 m_K^2\nonumber \\
& &+
\left(\frac{8}{3}-\frac{16}{27}\right) B_0^3 m_s \left( 2L_8 + H_2 \right)
\Bigg\} + \frac{\alpha_{em}}{12\pi}m_s B_0 \log\frac{\mu_0^2}{\Lambda^2}\ ,
\nonumber\\
\left. m_{K^+}^2\right|_{em}^{SD} &=&
\left. m_{\pi^+}^2\right|_{em}^{SD} + \left. m_{K^0}^2\right|_{em}^{SD}
+ \frac{\alpha_{em}\alpha_S}{\Lambda^2}
\Bigg\{\left(-8 +\frac{16}{9}\right) B_0^2 L_5 m_K^2
\nonumber\\
& & +\left(\frac{1}{2}-\frac{1}{18}\right) f^2 m_K^2
+\left( 96+ 0\right) B_0^3 m_s L_8 \Bigg\}\ .
\ea
The result in the limit of $m_s \to 0$ agrees with the result obtained earlier
in ref. \cite{BBG2}. The last term in the contribution to the $K^0$ mass
is the contribution from the electromagnetic current quark mass
renormalization. The term proportional to $\left(2L_8+H_2\right)$ is the term
which comes from the change in $\langle\overline{s}s\rangle$ due to $m_s$.
It agrees with the result for this change obtained in ref. \cite{GL} in the
limit of large $N_c$. Both these effects cancel out in the kaon mass
difference.
In eqs. (\ref{SD}) the first and second numbers inside the brackets
always refer to
the contribution from the box diagrams and the penguin diagrams,
respectively.
As could be expected the penguin diagrams
do contribute as soon as nonzero
quark masses are introduced.

It should be emphasized that this derivation is exact in the
leading order of the large $N_c$ expansion and to first non-leading order
in $m_s$. It does not suffer from the approximations used in the next section.

{\bf 3) Long distance :}
The long distance part of the integration over photon momenta requires
knowledge of the photon couplings to pseudoscalars and its $SU(3)$ breaking.
In the leading $N_c$ limit the $m_s$ dependence of this coupling only
shows up at $O(p^6)$ in the chiral counting. We will neglect these effects.
Similarly the effects of vector mesons in the chiral limit could be included
along the lines of ref. \cite{BBG2} but that method requires knowledge
of the $SU(3)$ breaking of the axial vector mesons. Using the hidden gauge
method to describe the vectors \cite{bando}
only the vector mesons could be included in a simple
fashion. This method
however seems to have some problems in the high energy behaviour. In the
case of the $\pi^+-\pi^0$ mass difference the quadratic divergence is only
cancelled for values of $a=1$ and not $a=2$ as
the best phenomenological
fit requires\cite{BG2}.
As such it does not seem useful to include in a first
estimate. We will only use here the lowest order chiral Lagrangian to describe
the photon-pseudoscalar couplings. This amounts to treating the pseudoscalars
as pointlike so the only effect comes from the internal and external mass
propagating in the diagrams.
There are two diagrams that contribute. They are shown in fig. \ref{fig2}.
There is no contribution to the $\pi^0$ or the $K^0$ mass from this
sort of diagrams.
The contribution is given by
\ba
\label{LD}
\left. m_{\pi^+}^2\right|_{em}^{LD} &=& \frac{3\alpha_{em}\Lambda^2}{4\pi}
\ ,\nonumber\\
\left. m_{K^+}^2\right|_{em}^{LD} &=&
e^2 i\int \frac{d^4 q}{(2\pi)^4}\left[
\frac{3}{q^2}-\frac{4m_K^2 + 2p\cdot q}{q^2\left(q^2 + 2 p\cdot q\right)}
\right]\ .
\ea
Here the integration ove $q$ has to be performed by rotating to Euclidean
momenta and keeping $q_E^2 \le \Lambda^2$. $p$ is the momentum of the kaon
with $p^2 = m_K^2$.
The first term in the integrand is equal to the contribution to the pion mass.
As can be seen the difference vanishes for $m_s \to 0$
as follows from Dashen's theorem. The correction to the kaon mass
is less divergent than the basic contribution.
For large value of $\Lambda$
the correction is logarithmically divergent. This already makes for a somewhat
smoother matching than is obtained for the basic result.

{\bf 4) Numerical results :}
In order to calculate the size of the corrections we need to know
the values of the higher order coefficients in the chiral Lagrangian.
These have been determined from measured quantities in ref. \cite{GL}.
I have used as values $L_8 = 0.0009$, $L_5 = 0.0014$, $f = 132~MeV$
a value of $m_K = 495~MeV$ and the value for $B_0$ derived
using the lowest order formula
$B_0 = m_K^2 / m_s(\Lambda)$ with a strange quark mass
of $150~MeV$ at a scale $\mu_0 = 1~GeV$ and a QCD coupling corresponding
to $\Lambda_{QCD} = 200~MeV$.
In fig. \ref{fig3} I have plotted the long distance contribution
to the charged pion mass squared and the extra correction needed
for the kaon mass squared difference.
As can be seen the corrections are sizable here.
Also plotted are the short distance
contributions to the pion mass squared and to the correction needed to
obtain the difference of the squared kaon masses.
The corrections are large and positive.
I have not plotted
the separate mass squareds of the kaons because they depend on the unknown
constant $H_2$ and there are gluonic corrections to $S_{eff}^1$ which
were not included so far and are also of order $\alpha_{em}\alpha_S m_s$.
In fig. \ref{fig4} the sums are shown. The top curve is the
mass difference in the chiral limit or $\dem{\pi}
= m_{\pi^+}^2 - m_{\pi^0}^2$ and the lower curve
is what should be added to $\dem{\pi}$ to obtain $
\dem{K}=m_{K^+}^2 - m_{K^0}^2$.
The experimental mass difference for the charged pion is given by about
$1.2\cdot 10^{-3}GeV^{2}$ so a reasonably good agreement is obtained there.

For the values of input given above if we choose the values obtained at
the place of minimum sensitivity to the matching scale we
obtain a $\dem{\pi}  = 1.27\cdot10^{-3}~GeV^{2}$ at a scale of $610~MeV$.
We obtain a value of $\dem{K}-\dem{\pi}
= 1.26\cdot 10^{-3}~GeV^{2}$ at $810~MeV$
and a combined result of $\dem{K} = 2.58\cdot 10^{-3}~GeV^{2}$
at 650~$MeV$. As can be seen
the corrections to Dashen's theorem are substantial and of order up to 100\%.
Abou one third from this correction came from the short distance contribution.
The result is somewhat sensitive to the value of $L_8$ which is rather
poorly known. Diminishing $L_8$ by its estimated error, using a value
of $L_8 = 0.0004$ instead we obtain
$\dem{K} = 2.31\cdot 10^{-3}~GeV^{2}$ at 600 $MeV$ and
$\dem{K}-\dem{\pi} = 1.04\cdot 10^{-3}~GeV^{2}$ at 570 $MeV$.
We can also include the corrections in the determination of the
constant $B_0$ from the lowest order relation used here.
This in fact absorbs the term proportional to $L_8$ in
$\dem{K}$. However the difference $\dem{K}-\dem{\pi}$ is only affected by this
at higher order. We would then in order to reproduce the observed pion
mass difference need to use a smaller value for the strange quark mass.
A reasonably conservative estimate of the uncertainty
involved then seems to be
\be
\left( m_{K^+}^2 - m_{K^0}^2 -m_{\pi^+}^2 + m_{\pi^0}^2 \right)_{em}
= (1.3 \pm 0.4)\cdot 10^{-3}~GeV^{2}\ .
\ee
These values correspond to
\be
\left. m_{\pi^+}\right|_{em} = 4.7~MeV
{}~~~~{\rm and}~~~~\left(m_{K^+} - m_{K^0}\right)_{em} = 2.6~MeV  \ .
\ee
The latter value should be compared to $1.3~MeV$ which is the value
obtained using Dashen's theorem. The effect of this is to increase
the value of $m_d - m_u$.

{\bf 5) Comparison with Ref. \cite{DHW} :}
In the recent preprint by Donoghue et al.\cite{DHW} the same problem was
considered in a somewhat different framework. I have calculated the
electromagnetic contribution to the mass within the framwework of QCD at
short-distances and used the model-independent part for the long distances.
In \cite{DHW} an approach similar to the one by Das et al. \cite{das}
was used. Here the long distance part is evaluated by calculating within
a simple model for the vector and axial-vector mesons where the
parameters are chosen such that the $SU(3)$ breaking effects in the
masses and the couplings are coupled in such a way as to allow the
extrapolation of $\Lambda\rightarrow\infty$. I.e. the couplings and
masses even in the presence of nonzero quark masses are chosen so
they still satisfy the first two Weinberg sum rules. The short-distance
part is then assumed to be modelled sufficiently well by this approximation.
Numerically the two approaches agree rather well. We also agree on the
long-distance part, eq. \ref{LD}, when the vector and axial-vector mesons
are removed from the expressions in Ref. \cite{DHW}.
In this reference the implications for the quark masses are also discussed
more extensively.

{\bf 6) Conclusions :} We have obtained within the framework of the $1/N_c$
expansion a sizeable correction to Dashen's theorem. This correction should
be subtracted from the numerator in eq. (\ref{msmd}) in order to
obtain a more accurate value of the quark mass ratios.

The short distance contribution was derived exactly in the limit of
large $N_c$ in terms of the parameters of the chiral Lagrangian.
The long distance contribution was evaluated in the approximation of
point-like pseudoscalars. A more accurate description would require an
understanding of $SU(3)$ breaking effects at intermediate length scales.
It has been shown in ref. \cite{ENJL} and references therein,
that the extended Nambu-Jona-Lasinio model describes the values of the
low-energy parameters well and also gives a reasonable description of
two-point functions at intermediate values of momenta. The estimate
of the long-distance correction to Dashen's theorem in that model is in
progress.

\section*{Acknowledgements}
I would like to thank H.~Leutwyler for reminding me of this problem.

\listoffigures
\newpage
\begin{figure}
\vspace{9cm}
\caption{The three types of short-distance contributions included.
(a) Corrections to the current quark mass, (b) Penguin diagrams, (c) Box
diagrams. The curly line is a photon, the full lines are quarks and the
dashed lines are gluons.}
\label{fig1}
\end{figure}
\begin{figure}
\vspace{10cm}
\caption{The long distance type of contributions. The curly line is a photon
and the full line is a $\pi^+$ or $K^+$.}
\label{fig2}
\end{figure}
\begin{figure}
\setlength{\unitlength}{0.240900pt}
\ifx\plotpoint\undefined\newsavebox{\plotpoint}\fi
\sbox{\plotpoint}{\rule[-0.175pt]{0.350pt}{0.350pt}}%
\begin{picture}(1500,900)(0,0)
\tenrm
\sbox{\plotpoint}{\rule[-0.175pt]{0.350pt}{0.350pt}}%
\put(264,158){\rule[-0.175pt]{282.335pt}{0.350pt}}
\put(264,158){\rule[-0.175pt]{4.818pt}{0.350pt}}
\put(242,158){\makebox(0,0)[r]{0}}
\put(1416,158){\rule[-0.175pt]{4.818pt}{0.350pt}}
\put(264,228){\rule[-0.175pt]{4.818pt}{0.350pt}}
\put(242,228){\makebox(0,0)[r]{2.0e-4}}
\put(1416,228){\rule[-0.175pt]{4.818pt}{0.350pt}}
\put(264,298){\rule[-0.175pt]{4.818pt}{0.350pt}}
\put(242,298){\makebox(0,0)[r]{4.0e-4}}
\put(1416,298){\rule[-0.175pt]{4.818pt}{0.350pt}}
\put(264,368){\rule[-0.175pt]{4.818pt}{0.350pt}}
\put(242,368){\makebox(0,0)[r]{6.0e-4}}
\put(1416,368){\rule[-0.175pt]{4.818pt}{0.350pt}}
\put(264,438){\rule[-0.175pt]{4.818pt}{0.350pt}}
\put(242,438){\makebox(0,0)[r]{8.0e-4}}
\put(1416,438){\rule[-0.175pt]{4.818pt}{0.350pt}}
\put(264,507){\rule[-0.175pt]{4.818pt}{0.350pt}}
\put(242,507){\makebox(0,0)[r]{1.0e-3}}
\put(1416,507){\rule[-0.175pt]{4.818pt}{0.350pt}}
\put(264,577){\rule[-0.175pt]{4.818pt}{0.350pt}}
\put(242,577){\makebox(0,0)[r]{1.2e-3}}
\put(1416,577){\rule[-0.175pt]{4.818pt}{0.350pt}}
\put(264,647){\rule[-0.175pt]{4.818pt}{0.350pt}}
\put(242,647){\makebox(0,0)[r]{1.4e-3}}
\put(1416,647){\rule[-0.175pt]{4.818pt}{0.350pt}}
\put(264,717){\rule[-0.175pt]{4.818pt}{0.350pt}}
\put(242,717){\makebox(0,0)[r]{1.6e-3}}
\put(1416,717){\rule[-0.175pt]{4.818pt}{0.350pt}}
\put(264,787){\rule[-0.175pt]{4.818pt}{0.350pt}}
\put(242,787){\makebox(0,0)[r]{1.8e-3}}
\put(1416,787){\rule[-0.175pt]{4.818pt}{0.350pt}}
\put(264,158){\rule[-0.175pt]{0.350pt}{4.818pt}}
\put(264,113){\makebox(0,0){0.5}}
\put(264,767){\rule[-0.175pt]{0.350pt}{4.818pt}}
\put(381,158){\rule[-0.175pt]{0.350pt}{4.818pt}}
\put(381,113){\makebox(0,0){0.55}}
\put(381,767){\rule[-0.175pt]{0.350pt}{4.818pt}}
\put(498,158){\rule[-0.175pt]{0.350pt}{4.818pt}}
\put(498,113){\makebox(0,0){0.6}}
\put(498,767){\rule[-0.175pt]{0.350pt}{4.818pt}}
\put(616,158){\rule[-0.175pt]{0.350pt}{4.818pt}}
\put(616,113){\makebox(0,0){0.65}}
\put(616,767){\rule[-0.175pt]{0.350pt}{4.818pt}}
\put(733,158){\rule[-0.175pt]{0.350pt}{4.818pt}}
\put(733,113){\makebox(0,0){0.7}}
\put(733,767){\rule[-0.175pt]{0.350pt}{4.818pt}}
\put(850,158){\rule[-0.175pt]{0.350pt}{4.818pt}}
\put(850,113){\makebox(0,0){0.75}}
\put(850,767){\rule[-0.175pt]{0.350pt}{4.818pt}}
\put(967,158){\rule[-0.175pt]{0.350pt}{4.818pt}}
\put(967,113){\makebox(0,0){0.8}}
\put(967,767){\rule[-0.175pt]{0.350pt}{4.818pt}}
\put(1084,158){\rule[-0.175pt]{0.350pt}{4.818pt}}
\put(1084,113){\makebox(0,0){0.85}}
\put(1084,767){\rule[-0.175pt]{0.350pt}{4.818pt}}
\put(1202,158){\rule[-0.175pt]{0.350pt}{4.818pt}}
\put(1202,113){\makebox(0,0){0.9}}
\put(1202,767){\rule[-0.175pt]{0.350pt}{4.818pt}}
\put(1319,158){\rule[-0.175pt]{0.350pt}{4.818pt}}
\put(1319,113){\makebox(0,0){0.95}}
\put(1319,767){\rule[-0.175pt]{0.350pt}{4.818pt}}
\put(1436,158){\rule[-0.175pt]{0.350pt}{4.818pt}}
\put(1436,113){\makebox(0,0){1}}
\put(1436,767){\rule[-0.175pt]{0.350pt}{4.818pt}}
\put(264,158){\rule[-0.175pt]{282.335pt}{0.350pt}}
\put(1436,158){\rule[-0.175pt]{0.350pt}{151.526pt}}
\put(264,787){\rule[-0.175pt]{282.335pt}{0.350pt}}
\put(45,472){\makebox(0,0)[l]{\shortstack{$GeV^2$}}}
\put(850,68){\makebox(0,0){$\Lambda~GeV$}}
\put(264,158){\rule[-0.175pt]{0.350pt}{151.526pt}}
\put(1306,722){\makebox(0,0)[r]{$m_\pi^2 ~LD$}}
\put(1328,722){\rule[-0.175pt]{15.899pt}{0.350pt}}
\put(264,310){\usebox{\plotpoint}}
\put(264,310){\rule[-0.175pt]{0.923pt}{0.350pt}}
\put(267,311){\rule[-0.175pt]{0.923pt}{0.350pt}}
\put(271,312){\rule[-0.175pt]{0.923pt}{0.350pt}}
\put(275,313){\rule[-0.175pt]{0.923pt}{0.350pt}}
\put(279,314){\rule[-0.175pt]{0.923pt}{0.350pt}}
\put(283,315){\rule[-0.175pt]{0.923pt}{0.350pt}}
\put(287,316){\rule[-0.175pt]{0.826pt}{0.350pt}}
\put(290,317){\rule[-0.175pt]{0.826pt}{0.350pt}}
\put(293,318){\rule[-0.175pt]{0.826pt}{0.350pt}}
\put(297,319){\rule[-0.175pt]{0.826pt}{0.350pt}}
\put(300,320){\rule[-0.175pt]{0.826pt}{0.350pt}}
\put(304,321){\rule[-0.175pt]{0.826pt}{0.350pt}}
\put(307,322){\rule[-0.175pt]{0.826pt}{0.350pt}}
\put(310,323){\rule[-0.175pt]{0.923pt}{0.350pt}}
\put(314,324){\rule[-0.175pt]{0.923pt}{0.350pt}}
\put(318,325){\rule[-0.175pt]{0.923pt}{0.350pt}}
\put(322,326){\rule[-0.175pt]{0.923pt}{0.350pt}}
\put(326,327){\rule[-0.175pt]{0.923pt}{0.350pt}}
\put(330,328){\rule[-0.175pt]{0.923pt}{0.350pt}}
\put(334,329){\rule[-0.175pt]{0.826pt}{0.350pt}}
\put(337,330){\rule[-0.175pt]{0.826pt}{0.350pt}}
\put(340,331){\rule[-0.175pt]{0.826pt}{0.350pt}}
\put(344,332){\rule[-0.175pt]{0.826pt}{0.350pt}}
\put(347,333){\rule[-0.175pt]{0.826pt}{0.350pt}}
\put(351,334){\rule[-0.175pt]{0.826pt}{0.350pt}}
\put(354,335){\rule[-0.175pt]{0.826pt}{0.350pt}}
\put(357,336){\rule[-0.175pt]{0.923pt}{0.350pt}}
\put(361,337){\rule[-0.175pt]{0.923pt}{0.350pt}}
\put(365,338){\rule[-0.175pt]{0.923pt}{0.350pt}}
\put(369,339){\rule[-0.175pt]{0.923pt}{0.350pt}}
\put(373,340){\rule[-0.175pt]{0.923pt}{0.350pt}}
\put(377,341){\rule[-0.175pt]{0.923pt}{0.350pt}}
\put(381,342){\rule[-0.175pt]{0.826pt}{0.350pt}}
\put(384,343){\rule[-0.175pt]{0.826pt}{0.350pt}}
\put(387,344){\rule[-0.175pt]{0.826pt}{0.350pt}}
\put(391,345){\rule[-0.175pt]{0.826pt}{0.350pt}}
\put(394,346){\rule[-0.175pt]{0.826pt}{0.350pt}}
\put(398,347){\rule[-0.175pt]{0.826pt}{0.350pt}}
\put(401,348){\rule[-0.175pt]{0.826pt}{0.350pt}}
\put(404,349){\rule[-0.175pt]{0.792pt}{0.350pt}}
\put(408,350){\rule[-0.175pt]{0.792pt}{0.350pt}}
\put(411,351){\rule[-0.175pt]{0.792pt}{0.350pt}}
\put(414,352){\rule[-0.175pt]{0.792pt}{0.350pt}}
\put(418,353){\rule[-0.175pt]{0.792pt}{0.350pt}}
\put(421,354){\rule[-0.175pt]{0.792pt}{0.350pt}}
\put(424,355){\rule[-0.175pt]{0.792pt}{0.350pt}}
\put(427,356){\rule[-0.175pt]{0.826pt}{0.350pt}}
\put(431,357){\rule[-0.175pt]{0.826pt}{0.350pt}}
\put(434,358){\rule[-0.175pt]{0.826pt}{0.350pt}}
\put(438,359){\rule[-0.175pt]{0.826pt}{0.350pt}}
\put(441,360){\rule[-0.175pt]{0.826pt}{0.350pt}}
\put(445,361){\rule[-0.175pt]{0.826pt}{0.350pt}}
\put(448,362){\rule[-0.175pt]{0.826pt}{0.350pt}}
\put(451,363){\rule[-0.175pt]{0.792pt}{0.350pt}}
\put(455,364){\rule[-0.175pt]{0.792pt}{0.350pt}}
\put(458,365){\rule[-0.175pt]{0.792pt}{0.350pt}}
\put(461,366){\rule[-0.175pt]{0.792pt}{0.350pt}}
\put(465,367){\rule[-0.175pt]{0.792pt}{0.350pt}}
\put(468,368){\rule[-0.175pt]{0.792pt}{0.350pt}}
\put(471,369){\rule[-0.175pt]{0.792pt}{0.350pt}}
\put(474,370){\rule[-0.175pt]{0.792pt}{0.350pt}}
\put(478,371){\rule[-0.175pt]{0.792pt}{0.350pt}}
\put(481,372){\rule[-0.175pt]{0.792pt}{0.350pt}}
\put(484,373){\rule[-0.175pt]{0.792pt}{0.350pt}}
\put(488,374){\rule[-0.175pt]{0.792pt}{0.350pt}}
\put(491,375){\rule[-0.175pt]{0.792pt}{0.350pt}}
\put(494,376){\rule[-0.175pt]{0.792pt}{0.350pt}}
\put(497,377){\rule[-0.175pt]{0.723pt}{0.350pt}}
\put(501,378){\rule[-0.175pt]{0.723pt}{0.350pt}}
\put(504,379){\rule[-0.175pt]{0.723pt}{0.350pt}}
\put(507,380){\rule[-0.175pt]{0.723pt}{0.350pt}}
\put(510,381){\rule[-0.175pt]{0.723pt}{0.350pt}}
\put(513,382){\rule[-0.175pt]{0.723pt}{0.350pt}}
\put(516,383){\rule[-0.175pt]{0.723pt}{0.350pt}}
\put(519,384){\rule[-0.175pt]{0.723pt}{0.350pt}}
\put(522,385){\rule[-0.175pt]{0.792pt}{0.350pt}}
\put(525,386){\rule[-0.175pt]{0.792pt}{0.350pt}}
\put(528,387){\rule[-0.175pt]{0.792pt}{0.350pt}}
\put(531,388){\rule[-0.175pt]{0.792pt}{0.350pt}}
\put(535,389){\rule[-0.175pt]{0.792pt}{0.350pt}}
\put(538,390){\rule[-0.175pt]{0.792pt}{0.350pt}}
\put(541,391){\rule[-0.175pt]{0.792pt}{0.350pt}}
\put(544,392){\rule[-0.175pt]{0.723pt}{0.350pt}}
\put(548,393){\rule[-0.175pt]{0.723pt}{0.350pt}}
\put(551,394){\rule[-0.175pt]{0.723pt}{0.350pt}}
\put(554,395){\rule[-0.175pt]{0.723pt}{0.350pt}}
\put(557,396){\rule[-0.175pt]{0.723pt}{0.350pt}}
\put(560,397){\rule[-0.175pt]{0.723pt}{0.350pt}}
\put(563,398){\rule[-0.175pt]{0.723pt}{0.350pt}}
\put(566,399){\rule[-0.175pt]{0.723pt}{0.350pt}}
\put(569,400){\rule[-0.175pt]{0.792pt}{0.350pt}}
\put(572,401){\rule[-0.175pt]{0.792pt}{0.350pt}}
\put(575,402){\rule[-0.175pt]{0.792pt}{0.350pt}}
\put(578,403){\rule[-0.175pt]{0.792pt}{0.350pt}}
\put(582,404){\rule[-0.175pt]{0.792pt}{0.350pt}}
\put(585,405){\rule[-0.175pt]{0.792pt}{0.350pt}}
\put(588,406){\rule[-0.175pt]{0.792pt}{0.350pt}}
\put(591,407){\rule[-0.175pt]{0.723pt}{0.350pt}}
\put(595,408){\rule[-0.175pt]{0.723pt}{0.350pt}}
\put(598,409){\rule[-0.175pt]{0.723pt}{0.350pt}}
\put(601,410){\rule[-0.175pt]{0.723pt}{0.350pt}}
\put(604,411){\rule[-0.175pt]{0.723pt}{0.350pt}}
\put(607,412){\rule[-0.175pt]{0.723pt}{0.350pt}}
\put(610,413){\rule[-0.175pt]{0.723pt}{0.350pt}}
\put(613,414){\rule[-0.175pt]{0.723pt}{0.350pt}}
\put(616,415){\rule[-0.175pt]{0.693pt}{0.350pt}}
\put(618,416){\rule[-0.175pt]{0.693pt}{0.350pt}}
\put(621,417){\rule[-0.175pt]{0.693pt}{0.350pt}}
\put(624,418){\rule[-0.175pt]{0.693pt}{0.350pt}}
\put(627,419){\rule[-0.175pt]{0.693pt}{0.350pt}}
\put(630,420){\rule[-0.175pt]{0.693pt}{0.350pt}}
\put(633,421){\rule[-0.175pt]{0.693pt}{0.350pt}}
\put(636,422){\rule[-0.175pt]{0.693pt}{0.350pt}}
\put(639,423){\rule[-0.175pt]{0.693pt}{0.350pt}}
\put(641,424){\rule[-0.175pt]{0.693pt}{0.350pt}}
\put(644,425){\rule[-0.175pt]{0.693pt}{0.350pt}}
\put(647,426){\rule[-0.175pt]{0.693pt}{0.350pt}}
\put(650,427){\rule[-0.175pt]{0.693pt}{0.350pt}}
\put(653,428){\rule[-0.175pt]{0.693pt}{0.350pt}}
\put(656,429){\rule[-0.175pt]{0.693pt}{0.350pt}}
\put(659,430){\rule[-0.175pt]{0.693pt}{0.350pt}}
\put(662,431){\rule[-0.175pt]{0.642pt}{0.350pt}}
\put(664,432){\rule[-0.175pt]{0.642pt}{0.350pt}}
\put(667,433){\rule[-0.175pt]{0.642pt}{0.350pt}}
\put(670,434){\rule[-0.175pt]{0.642pt}{0.350pt}}
\put(672,435){\rule[-0.175pt]{0.642pt}{0.350pt}}
\put(675,436){\rule[-0.175pt]{0.642pt}{0.350pt}}
\put(678,437){\rule[-0.175pt]{0.642pt}{0.350pt}}
\put(680,438){\rule[-0.175pt]{0.642pt}{0.350pt}}
\put(683,439){\rule[-0.175pt]{0.642pt}{0.350pt}}
\put(686,440){\rule[-0.175pt]{0.693pt}{0.350pt}}
\put(688,441){\rule[-0.175pt]{0.693pt}{0.350pt}}
\put(691,442){\rule[-0.175pt]{0.693pt}{0.350pt}}
\put(694,443){\rule[-0.175pt]{0.693pt}{0.350pt}}
\put(697,444){\rule[-0.175pt]{0.693pt}{0.350pt}}
\put(700,445){\rule[-0.175pt]{0.693pt}{0.350pt}}
\put(703,446){\rule[-0.175pt]{0.693pt}{0.350pt}}
\put(706,447){\rule[-0.175pt]{0.693pt}{0.350pt}}
\put(709,448){\rule[-0.175pt]{0.723pt}{0.350pt}}
\put(712,449){\rule[-0.175pt]{0.723pt}{0.350pt}}
\put(715,450){\rule[-0.175pt]{0.723pt}{0.350pt}}
\put(718,451){\rule[-0.175pt]{0.723pt}{0.350pt}}
\put(721,452){\rule[-0.175pt]{0.723pt}{0.350pt}}
\put(724,453){\rule[-0.175pt]{0.723pt}{0.350pt}}
\put(727,454){\rule[-0.175pt]{0.723pt}{0.350pt}}
\put(730,455){\rule[-0.175pt]{0.723pt}{0.350pt}}
\put(733,456){\rule[-0.175pt]{0.616pt}{0.350pt}}
\put(735,457){\rule[-0.175pt]{0.616pt}{0.350pt}}
\put(738,458){\rule[-0.175pt]{0.616pt}{0.350pt}}
\put(740,459){\rule[-0.175pt]{0.616pt}{0.350pt}}
\put(743,460){\rule[-0.175pt]{0.616pt}{0.350pt}}
\put(745,461){\rule[-0.175pt]{0.616pt}{0.350pt}}
\put(748,462){\rule[-0.175pt]{0.616pt}{0.350pt}}
\put(750,463){\rule[-0.175pt]{0.616pt}{0.350pt}}
\put(753,464){\rule[-0.175pt]{0.616pt}{0.350pt}}
\put(755,465){\rule[-0.175pt]{0.642pt}{0.350pt}}
\put(758,466){\rule[-0.175pt]{0.642pt}{0.350pt}}
\put(761,467){\rule[-0.175pt]{0.642pt}{0.350pt}}
\put(764,468){\rule[-0.175pt]{0.642pt}{0.350pt}}
\put(766,469){\rule[-0.175pt]{0.642pt}{0.350pt}}
\put(769,470){\rule[-0.175pt]{0.642pt}{0.350pt}}
\put(772,471){\rule[-0.175pt]{0.642pt}{0.350pt}}
\put(774,472){\rule[-0.175pt]{0.642pt}{0.350pt}}
\put(777,473){\rule[-0.175pt]{0.642pt}{0.350pt}}
\put(780,474){\rule[-0.175pt]{0.693pt}{0.350pt}}
\put(782,475){\rule[-0.175pt]{0.693pt}{0.350pt}}
\put(785,476){\rule[-0.175pt]{0.693pt}{0.350pt}}
\put(788,477){\rule[-0.175pt]{0.693pt}{0.350pt}}
\put(791,478){\rule[-0.175pt]{0.693pt}{0.350pt}}
\put(794,479){\rule[-0.175pt]{0.693pt}{0.350pt}}
\put(797,480){\rule[-0.175pt]{0.693pt}{0.350pt}}
\put(800,481){\rule[-0.175pt]{0.693pt}{0.350pt}}
\put(803,482){\rule[-0.175pt]{0.642pt}{0.350pt}}
\put(805,483){\rule[-0.175pt]{0.642pt}{0.350pt}}
\put(808,484){\rule[-0.175pt]{0.642pt}{0.350pt}}
\put(811,485){\rule[-0.175pt]{0.642pt}{0.350pt}}
\put(813,486){\rule[-0.175pt]{0.642pt}{0.350pt}}
\put(816,487){\rule[-0.175pt]{0.642pt}{0.350pt}}
\put(819,488){\rule[-0.175pt]{0.642pt}{0.350pt}}
\put(821,489){\rule[-0.175pt]{0.642pt}{0.350pt}}
\put(824,490){\rule[-0.175pt]{0.642pt}{0.350pt}}
\put(827,491){\rule[-0.175pt]{0.554pt}{0.350pt}}
\put(829,492){\rule[-0.175pt]{0.554pt}{0.350pt}}
\put(831,493){\rule[-0.175pt]{0.554pt}{0.350pt}}
\put(833,494){\rule[-0.175pt]{0.554pt}{0.350pt}}
\put(836,495){\rule[-0.175pt]{0.554pt}{0.350pt}}
\put(838,496){\rule[-0.175pt]{0.554pt}{0.350pt}}
\put(840,497){\rule[-0.175pt]{0.554pt}{0.350pt}}
\put(843,498){\rule[-0.175pt]{0.554pt}{0.350pt}}
\put(845,499){\rule[-0.175pt]{0.554pt}{0.350pt}}
\put(847,500){\rule[-0.175pt]{0.554pt}{0.350pt}}
\put(849,501){\rule[-0.175pt]{0.616pt}{0.350pt}}
\put(852,502){\rule[-0.175pt]{0.616pt}{0.350pt}}
\put(855,503){\rule[-0.175pt]{0.616pt}{0.350pt}}
\put(857,504){\rule[-0.175pt]{0.616pt}{0.350pt}}
\put(860,505){\rule[-0.175pt]{0.616pt}{0.350pt}}
\put(862,506){\rule[-0.175pt]{0.616pt}{0.350pt}}
\put(865,507){\rule[-0.175pt]{0.616pt}{0.350pt}}
\put(867,508){\rule[-0.175pt]{0.616pt}{0.350pt}}
\put(870,509){\rule[-0.175pt]{0.616pt}{0.350pt}}
\put(872,510){\rule[-0.175pt]{0.642pt}{0.350pt}}
\put(875,511){\rule[-0.175pt]{0.642pt}{0.350pt}}
\put(878,512){\rule[-0.175pt]{0.642pt}{0.350pt}}
\put(881,513){\rule[-0.175pt]{0.642pt}{0.350pt}}
\put(883,514){\rule[-0.175pt]{0.642pt}{0.350pt}}
\put(886,515){\rule[-0.175pt]{0.642pt}{0.350pt}}
\put(889,516){\rule[-0.175pt]{0.642pt}{0.350pt}}
\put(891,517){\rule[-0.175pt]{0.642pt}{0.350pt}}
\put(894,518){\rule[-0.175pt]{0.642pt}{0.350pt}}
\put(897,519){\rule[-0.175pt]{0.616pt}{0.350pt}}
\put(899,520){\rule[-0.175pt]{0.616pt}{0.350pt}}
\put(902,521){\rule[-0.175pt]{0.616pt}{0.350pt}}
\put(904,522){\rule[-0.175pt]{0.616pt}{0.350pt}}
\put(907,523){\rule[-0.175pt]{0.616pt}{0.350pt}}
\put(909,524){\rule[-0.175pt]{0.616pt}{0.350pt}}
\put(912,525){\rule[-0.175pt]{0.616pt}{0.350pt}}
\put(914,526){\rule[-0.175pt]{0.616pt}{0.350pt}}
\put(917,527){\rule[-0.175pt]{0.616pt}{0.350pt}}
\put(919,528){\rule[-0.175pt]{0.578pt}{0.350pt}}
\put(922,529){\rule[-0.175pt]{0.578pt}{0.350pt}}
\put(924,530){\rule[-0.175pt]{0.578pt}{0.350pt}}
\put(927,531){\rule[-0.175pt]{0.578pt}{0.350pt}}
\put(929,532){\rule[-0.175pt]{0.578pt}{0.350pt}}
\put(932,533){\rule[-0.175pt]{0.578pt}{0.350pt}}
\put(934,534){\rule[-0.175pt]{0.578pt}{0.350pt}}
\put(936,535){\rule[-0.175pt]{0.578pt}{0.350pt}}
\put(939,536){\rule[-0.175pt]{0.578pt}{0.350pt}}
\put(941,537){\rule[-0.175pt]{0.578pt}{0.350pt}}
\put(944,538){\rule[-0.175pt]{0.554pt}{0.350pt}}
\put(946,539){\rule[-0.175pt]{0.554pt}{0.350pt}}
\put(948,540){\rule[-0.175pt]{0.554pt}{0.350pt}}
\put(950,541){\rule[-0.175pt]{0.554pt}{0.350pt}}
\put(953,542){\rule[-0.175pt]{0.554pt}{0.350pt}}
\put(955,543){\rule[-0.175pt]{0.554pt}{0.350pt}}
\put(957,544){\rule[-0.175pt]{0.554pt}{0.350pt}}
\put(960,545){\rule[-0.175pt]{0.554pt}{0.350pt}}
\put(962,546){\rule[-0.175pt]{0.554pt}{0.350pt}}
\put(964,547){\rule[-0.175pt]{0.554pt}{0.350pt}}
\put(966,548){\rule[-0.175pt]{0.578pt}{0.350pt}}
\put(969,549){\rule[-0.175pt]{0.578pt}{0.350pt}}
\put(971,550){\rule[-0.175pt]{0.578pt}{0.350pt}}
\put(974,551){\rule[-0.175pt]{0.578pt}{0.350pt}}
\put(976,552){\rule[-0.175pt]{0.578pt}{0.350pt}}
\put(979,553){\rule[-0.175pt]{0.578pt}{0.350pt}}
\put(981,554){\rule[-0.175pt]{0.578pt}{0.350pt}}
\put(983,555){\rule[-0.175pt]{0.578pt}{0.350pt}}
\put(986,556){\rule[-0.175pt]{0.578pt}{0.350pt}}
\put(988,557){\rule[-0.175pt]{0.578pt}{0.350pt}}
\put(991,558){\rule[-0.175pt]{0.616pt}{0.350pt}}
\put(993,559){\rule[-0.175pt]{0.616pt}{0.350pt}}
\put(996,560){\rule[-0.175pt]{0.616pt}{0.350pt}}
\put(998,561){\rule[-0.175pt]{0.616pt}{0.350pt}}
\put(1001,562){\rule[-0.175pt]{0.616pt}{0.350pt}}
\put(1003,563){\rule[-0.175pt]{0.616pt}{0.350pt}}
\put(1006,564){\rule[-0.175pt]{0.616pt}{0.350pt}}
\put(1008,565){\rule[-0.175pt]{0.616pt}{0.350pt}}
\put(1011,566){\rule[-0.175pt]{0.616pt}{0.350pt}}
\put(1013,567){\rule[-0.175pt]{0.578pt}{0.350pt}}
\put(1016,568){\rule[-0.175pt]{0.578pt}{0.350pt}}
\put(1018,569){\rule[-0.175pt]{0.578pt}{0.350pt}}
\put(1021,570){\rule[-0.175pt]{0.578pt}{0.350pt}}
\put(1023,571){\rule[-0.175pt]{0.578pt}{0.350pt}}
\put(1026,572){\rule[-0.175pt]{0.578pt}{0.350pt}}
\put(1028,573){\rule[-0.175pt]{0.578pt}{0.350pt}}
\put(1030,574){\rule[-0.175pt]{0.578pt}{0.350pt}}
\put(1033,575){\rule[-0.175pt]{0.578pt}{0.350pt}}
\put(1035,576){\rule[-0.175pt]{0.578pt}{0.350pt}}
\put(1038,577){\rule[-0.175pt]{0.504pt}{0.350pt}}
\put(1040,578){\rule[-0.175pt]{0.504pt}{0.350pt}}
\put(1042,579){\rule[-0.175pt]{0.504pt}{0.350pt}}
\put(1044,580){\rule[-0.175pt]{0.504pt}{0.350pt}}
\put(1046,581){\rule[-0.175pt]{0.504pt}{0.350pt}}
\put(1048,582){\rule[-0.175pt]{0.504pt}{0.350pt}}
\put(1050,583){\rule[-0.175pt]{0.504pt}{0.350pt}}
\put(1052,584){\rule[-0.175pt]{0.504pt}{0.350pt}}
\put(1054,585){\rule[-0.175pt]{0.504pt}{0.350pt}}
\put(1056,586){\rule[-0.175pt]{0.504pt}{0.350pt}}
\put(1058,587){\rule[-0.175pt]{0.504pt}{0.350pt}}
\put(1061,588){\rule[-0.175pt]{0.554pt}{0.350pt}}
\put(1063,589){\rule[-0.175pt]{0.554pt}{0.350pt}}
\put(1065,590){\rule[-0.175pt]{0.554pt}{0.350pt}}
\put(1067,591){\rule[-0.175pt]{0.554pt}{0.350pt}}
\put(1070,592){\rule[-0.175pt]{0.554pt}{0.350pt}}
\put(1072,593){\rule[-0.175pt]{0.554pt}{0.350pt}}
\put(1074,594){\rule[-0.175pt]{0.554pt}{0.350pt}}
\put(1077,595){\rule[-0.175pt]{0.554pt}{0.350pt}}
\put(1079,596){\rule[-0.175pt]{0.554pt}{0.350pt}}
\put(1081,597){\rule[-0.175pt]{0.554pt}{0.350pt}}
\put(1084,598){\rule[-0.175pt]{0.578pt}{0.350pt}}
\put(1086,599){\rule[-0.175pt]{0.578pt}{0.350pt}}
\put(1088,600){\rule[-0.175pt]{0.578pt}{0.350pt}}
\put(1091,601){\rule[-0.175pt]{0.578pt}{0.350pt}}
\put(1093,602){\rule[-0.175pt]{0.578pt}{0.350pt}}
\put(1096,603){\rule[-0.175pt]{0.578pt}{0.350pt}}
\put(1098,604){\rule[-0.175pt]{0.578pt}{0.350pt}}
\put(1100,605){\rule[-0.175pt]{0.578pt}{0.350pt}}
\put(1103,606){\rule[-0.175pt]{0.578pt}{0.350pt}}
\put(1105,607){\rule[-0.175pt]{0.578pt}{0.350pt}}
\put(1108,608){\rule[-0.175pt]{0.504pt}{0.350pt}}
\put(1110,609){\rule[-0.175pt]{0.504pt}{0.350pt}}
\put(1112,610){\rule[-0.175pt]{0.504pt}{0.350pt}}
\put(1114,611){\rule[-0.175pt]{0.504pt}{0.350pt}}
\put(1116,612){\rule[-0.175pt]{0.504pt}{0.350pt}}
\put(1118,613){\rule[-0.175pt]{0.504pt}{0.350pt}}
\put(1120,614){\rule[-0.175pt]{0.504pt}{0.350pt}}
\put(1122,615){\rule[-0.175pt]{0.504pt}{0.350pt}}
\put(1124,616){\rule[-0.175pt]{0.504pt}{0.350pt}}
\put(1126,617){\rule[-0.175pt]{0.504pt}{0.350pt}}
\put(1128,618){\rule[-0.175pt]{0.504pt}{0.350pt}}
\put(1131,619){\rule[-0.175pt]{0.526pt}{0.350pt}}
\put(1133,620){\rule[-0.175pt]{0.526pt}{0.350pt}}
\put(1135,621){\rule[-0.175pt]{0.526pt}{0.350pt}}
\put(1137,622){\rule[-0.175pt]{0.526pt}{0.350pt}}
\put(1139,623){\rule[-0.175pt]{0.526pt}{0.350pt}}
\put(1141,624){\rule[-0.175pt]{0.526pt}{0.350pt}}
\put(1144,625){\rule[-0.175pt]{0.526pt}{0.350pt}}
\put(1146,626){\rule[-0.175pt]{0.526pt}{0.350pt}}
\put(1148,627){\rule[-0.175pt]{0.526pt}{0.350pt}}
\put(1150,628){\rule[-0.175pt]{0.526pt}{0.350pt}}
\put(1152,629){\rule[-0.175pt]{0.526pt}{0.350pt}}
\put(1154,630){\rule[-0.175pt]{0.554pt}{0.350pt}}
\put(1157,631){\rule[-0.175pt]{0.554pt}{0.350pt}}
\put(1159,632){\rule[-0.175pt]{0.554pt}{0.350pt}}
\put(1161,633){\rule[-0.175pt]{0.554pt}{0.350pt}}
\put(1164,634){\rule[-0.175pt]{0.554pt}{0.350pt}}
\put(1166,635){\rule[-0.175pt]{0.554pt}{0.350pt}}
\put(1168,636){\rule[-0.175pt]{0.554pt}{0.350pt}}
\put(1171,637){\rule[-0.175pt]{0.554pt}{0.350pt}}
\put(1173,638){\rule[-0.175pt]{0.554pt}{0.350pt}}
\put(1175,639){\rule[-0.175pt]{0.554pt}{0.350pt}}
\put(1178,640){\rule[-0.175pt]{0.526pt}{0.350pt}}
\put(1180,641){\rule[-0.175pt]{0.526pt}{0.350pt}}
\put(1182,642){\rule[-0.175pt]{0.526pt}{0.350pt}}
\put(1184,643){\rule[-0.175pt]{0.526pt}{0.350pt}}
\put(1186,644){\rule[-0.175pt]{0.526pt}{0.350pt}}
\put(1188,645){\rule[-0.175pt]{0.526pt}{0.350pt}}
\put(1191,646){\rule[-0.175pt]{0.526pt}{0.350pt}}
\put(1193,647){\rule[-0.175pt]{0.526pt}{0.350pt}}
\put(1195,648){\rule[-0.175pt]{0.526pt}{0.350pt}}
\put(1197,649){\rule[-0.175pt]{0.526pt}{0.350pt}}
\put(1199,650){\rule[-0.175pt]{0.526pt}{0.350pt}}
\put(1201,651){\rule[-0.175pt]{0.504pt}{0.350pt}}
\put(1204,652){\rule[-0.175pt]{0.504pt}{0.350pt}}
\put(1206,653){\rule[-0.175pt]{0.504pt}{0.350pt}}
\put(1208,654){\rule[-0.175pt]{0.504pt}{0.350pt}}
\put(1210,655){\rule[-0.175pt]{0.504pt}{0.350pt}}
\put(1212,656){\rule[-0.175pt]{0.504pt}{0.350pt}}
\put(1214,657){\rule[-0.175pt]{0.504pt}{0.350pt}}
\put(1216,658){\rule[-0.175pt]{0.504pt}{0.350pt}}
\put(1218,659){\rule[-0.175pt]{0.504pt}{0.350pt}}
\put(1220,660){\rule[-0.175pt]{0.504pt}{0.350pt}}
\put(1222,661){\rule[-0.175pt]{0.504pt}{0.350pt}}
\put(1225,662){\rule[-0.175pt]{0.504pt}{0.350pt}}
\put(1227,663){\rule[-0.175pt]{0.504pt}{0.350pt}}
\put(1229,664){\rule[-0.175pt]{0.504pt}{0.350pt}}
\put(1231,665){\rule[-0.175pt]{0.504pt}{0.350pt}}
\put(1233,666){\rule[-0.175pt]{0.504pt}{0.350pt}}
\put(1235,667){\rule[-0.175pt]{0.504pt}{0.350pt}}
\put(1237,668){\rule[-0.175pt]{0.504pt}{0.350pt}}
\put(1239,669){\rule[-0.175pt]{0.504pt}{0.350pt}}
\put(1241,670){\rule[-0.175pt]{0.504pt}{0.350pt}}
\put(1243,671){\rule[-0.175pt]{0.504pt}{0.350pt}}
\put(1245,672){\rule[-0.175pt]{0.504pt}{0.350pt}}
\put(1248,673){\rule[-0.175pt]{0.482pt}{0.350pt}}
\put(1250,674){\rule[-0.175pt]{0.482pt}{0.350pt}}
\put(1252,675){\rule[-0.175pt]{0.482pt}{0.350pt}}
\put(1254,676){\rule[-0.175pt]{0.482pt}{0.350pt}}
\put(1256,677){\rule[-0.175pt]{0.482pt}{0.350pt}}
\put(1258,678){\rule[-0.175pt]{0.482pt}{0.350pt}}
\put(1260,679){\rule[-0.175pt]{0.482pt}{0.350pt}}
\put(1262,680){\rule[-0.175pt]{0.482pt}{0.350pt}}
\put(1264,681){\rule[-0.175pt]{0.482pt}{0.350pt}}
\put(1266,682){\rule[-0.175pt]{0.482pt}{0.350pt}}
\put(1268,683){\rule[-0.175pt]{0.482pt}{0.350pt}}
\put(1270,684){\rule[-0.175pt]{0.482pt}{0.350pt}}
\put(1272,685){\rule[-0.175pt]{0.504pt}{0.350pt}}
\put(1274,686){\rule[-0.175pt]{0.504pt}{0.350pt}}
\put(1276,687){\rule[-0.175pt]{0.504pt}{0.350pt}}
\put(1278,688){\rule[-0.175pt]{0.504pt}{0.350pt}}
\put(1280,689){\rule[-0.175pt]{0.504pt}{0.350pt}}
\put(1282,690){\rule[-0.175pt]{0.504pt}{0.350pt}}
\put(1284,691){\rule[-0.175pt]{0.504pt}{0.350pt}}
\put(1286,692){\rule[-0.175pt]{0.504pt}{0.350pt}}
\put(1288,693){\rule[-0.175pt]{0.504pt}{0.350pt}}
\put(1290,694){\rule[-0.175pt]{0.504pt}{0.350pt}}
\put(1292,695){\rule[-0.175pt]{0.504pt}{0.350pt}}
\put(1295,696){\rule[-0.175pt]{0.482pt}{0.350pt}}
\put(1297,697){\rule[-0.175pt]{0.482pt}{0.350pt}}
\put(1299,698){\rule[-0.175pt]{0.482pt}{0.350pt}}
\put(1301,699){\rule[-0.175pt]{0.482pt}{0.350pt}}
\put(1303,700){\rule[-0.175pt]{0.482pt}{0.350pt}}
\put(1305,701){\rule[-0.175pt]{0.482pt}{0.350pt}}
\put(1307,702){\rule[-0.175pt]{0.482pt}{0.350pt}}
\put(1309,703){\rule[-0.175pt]{0.482pt}{0.350pt}}
\put(1311,704){\rule[-0.175pt]{0.482pt}{0.350pt}}
\put(1313,705){\rule[-0.175pt]{0.482pt}{0.350pt}}
\put(1315,706){\rule[-0.175pt]{0.482pt}{0.350pt}}
\put(1317,707){\rule[-0.175pt]{0.482pt}{0.350pt}}
\put(1319,708){\rule[-0.175pt]{0.504pt}{0.350pt}}
\put(1321,709){\rule[-0.175pt]{0.504pt}{0.350pt}}
\put(1323,710){\rule[-0.175pt]{0.504pt}{0.350pt}}
\put(1325,711){\rule[-0.175pt]{0.504pt}{0.350pt}}
\put(1327,712){\rule[-0.175pt]{0.504pt}{0.350pt}}
\put(1329,713){\rule[-0.175pt]{0.504pt}{0.350pt}}
\put(1331,714){\rule[-0.175pt]{0.504pt}{0.350pt}}
\put(1333,715){\rule[-0.175pt]{0.504pt}{0.350pt}}
\put(1335,716){\rule[-0.175pt]{0.504pt}{0.350pt}}
\put(1337,717){\rule[-0.175pt]{0.504pt}{0.350pt}}
\put(1339,718){\rule[-0.175pt]{0.504pt}{0.350pt}}
\put(1342,719){\rule[-0.175pt]{0.482pt}{0.350pt}}
\put(1344,720){\rule[-0.175pt]{0.482pt}{0.350pt}}
\put(1346,721){\rule[-0.175pt]{0.482pt}{0.350pt}}
\put(1348,722){\rule[-0.175pt]{0.482pt}{0.350pt}}
\put(1350,723){\rule[-0.175pt]{0.482pt}{0.350pt}}
\put(1352,724){\rule[-0.175pt]{0.482pt}{0.350pt}}
\put(1354,725){\rule[-0.175pt]{0.482pt}{0.350pt}}
\put(1356,726){\rule[-0.175pt]{0.482pt}{0.350pt}}
\put(1358,727){\rule[-0.175pt]{0.482pt}{0.350pt}}
\put(1360,728){\rule[-0.175pt]{0.482pt}{0.350pt}}
\put(1362,729){\rule[-0.175pt]{0.482pt}{0.350pt}}
\put(1364,730){\rule[-0.175pt]{0.482pt}{0.350pt}}
\put(1366,731){\rule[-0.175pt]{0.462pt}{0.350pt}}
\put(1367,732){\rule[-0.175pt]{0.462pt}{0.350pt}}
\put(1369,733){\rule[-0.175pt]{0.462pt}{0.350pt}}
\put(1371,734){\rule[-0.175pt]{0.462pt}{0.350pt}}
\put(1373,735){\rule[-0.175pt]{0.462pt}{0.350pt}}
\put(1375,736){\rule[-0.175pt]{0.462pt}{0.350pt}}
\put(1377,737){\rule[-0.175pt]{0.462pt}{0.350pt}}
\put(1379,738){\rule[-0.175pt]{0.462pt}{0.350pt}}
\put(1381,739){\rule[-0.175pt]{0.462pt}{0.350pt}}
\put(1383,740){\rule[-0.175pt]{0.462pt}{0.350pt}}
\put(1385,741){\rule[-0.175pt]{0.462pt}{0.350pt}}
\put(1387,742){\rule[-0.175pt]{0.462pt}{0.350pt}}
\put(1388,743){\rule[-0.175pt]{0.482pt}{0.350pt}}
\put(1391,744){\rule[-0.175pt]{0.482pt}{0.350pt}}
\put(1393,745){\rule[-0.175pt]{0.482pt}{0.350pt}}
\put(1395,746){\rule[-0.175pt]{0.482pt}{0.350pt}}
\put(1397,747){\rule[-0.175pt]{0.482pt}{0.350pt}}
\put(1399,748){\rule[-0.175pt]{0.482pt}{0.350pt}}
\put(1401,749){\rule[-0.175pt]{0.482pt}{0.350pt}}
\put(1403,750){\rule[-0.175pt]{0.482pt}{0.350pt}}
\put(1405,751){\rule[-0.175pt]{0.482pt}{0.350pt}}
\put(1407,752){\rule[-0.175pt]{0.482pt}{0.350pt}}
\put(1409,753){\rule[-0.175pt]{0.482pt}{0.350pt}}
\put(1411,754){\rule[-0.175pt]{0.482pt}{0.350pt}}
\put(1413,755){\rule[-0.175pt]{0.462pt}{0.350pt}}
\put(1414,756){\rule[-0.175pt]{0.462pt}{0.350pt}}
\put(1416,757){\rule[-0.175pt]{0.462pt}{0.350pt}}
\put(1418,758){\rule[-0.175pt]{0.462pt}{0.350pt}}
\put(1420,759){\rule[-0.175pt]{0.462pt}{0.350pt}}
\put(1422,760){\rule[-0.175pt]{0.462pt}{0.350pt}}
\put(1424,761){\rule[-0.175pt]{0.462pt}{0.350pt}}
\put(1426,762){\rule[-0.175pt]{0.462pt}{0.350pt}}
\put(1428,763){\rule[-0.175pt]{0.462pt}{0.350pt}}
\put(1430,764){\rule[-0.175pt]{0.462pt}{0.350pt}}
\put(1432,765){\rule[-0.175pt]{0.462pt}{0.350pt}}
\put(1434,766){\rule[-0.175pt]{0.462pt}{0.350pt}}
\put(1435,767){\usebox{\plotpoint}}
\sbox{\plotpoint}{\rule[-0.350pt]{0.700pt}{0.700pt}}%
\put(1306,677){\makebox(0,0)[r]{$\Delta D~LD$}}
\put(1328,677){\rule[-0.350pt]{15.899pt}{0.700pt}}
\put(264,421){\usebox{\plotpoint}}
\put(264,421){\rule[-0.350pt]{1.847pt}{0.700pt}}
\put(271,422){\rule[-0.350pt]{1.847pt}{0.700pt}}
\put(279,423){\rule[-0.350pt]{1.847pt}{0.700pt}}
\put(286,424){\rule[-0.350pt]{1.445pt}{0.700pt}}
\put(293,425){\rule[-0.350pt]{1.445pt}{0.700pt}}
\put(299,426){\rule[-0.350pt]{1.445pt}{0.700pt}}
\put(305,427){\rule[-0.350pt]{1.445pt}{0.700pt}}
\put(311,428){\rule[-0.350pt]{1.847pt}{0.700pt}}
\put(318,429){\rule[-0.350pt]{1.847pt}{0.700pt}}
\put(326,430){\rule[-0.350pt]{1.847pt}{0.700pt}}
\put(333,431){\rule[-0.350pt]{1.927pt}{0.700pt}}
\put(342,432){\rule[-0.350pt]{1.927pt}{0.700pt}}
\put(350,433){\rule[-0.350pt]{1.927pt}{0.700pt}}
\put(358,434){\rule[-0.350pt]{1.847pt}{0.700pt}}
\put(365,435){\rule[-0.350pt]{1.847pt}{0.700pt}}
\put(373,436){\rule[-0.350pt]{1.847pt}{0.700pt}}
\put(380,437){\rule[-0.350pt]{1.445pt}{0.700pt}}
\put(387,438){\rule[-0.350pt]{1.445pt}{0.700pt}}
\put(393,439){\rule[-0.350pt]{1.445pt}{0.700pt}}
\put(399,440){\rule[-0.350pt]{1.445pt}{0.700pt}}
\put(405,441){\rule[-0.350pt]{1.847pt}{0.700pt}}
\put(412,442){\rule[-0.350pt]{1.847pt}{0.700pt}}
\put(420,443){\rule[-0.350pt]{1.847pt}{0.700pt}}
\put(427,444){\rule[-0.350pt]{1.927pt}{0.700pt}}
\put(436,445){\rule[-0.350pt]{1.927pt}{0.700pt}}
\put(444,446){\rule[-0.350pt]{1.927pt}{0.700pt}}
\put(452,447){\rule[-0.350pt]{1.847pt}{0.700pt}}
\put(459,448){\rule[-0.350pt]{1.847pt}{0.700pt}}
\put(467,449){\rule[-0.350pt]{1.847pt}{0.700pt}}
\put(474,450){\rule[-0.350pt]{1.847pt}{0.700pt}}
\put(482,451){\rule[-0.350pt]{1.847pt}{0.700pt}}
\put(490,452){\rule[-0.350pt]{1.847pt}{0.700pt}}
\put(497,453){\rule[-0.350pt]{1.927pt}{0.700pt}}
\put(506,454){\rule[-0.350pt]{1.927pt}{0.700pt}}
\put(514,455){\rule[-0.350pt]{1.927pt}{0.700pt}}
\put(522,456){\rule[-0.350pt]{1.847pt}{0.700pt}}
\put(529,457){\rule[-0.350pt]{1.847pt}{0.700pt}}
\put(537,458){\rule[-0.350pt]{1.847pt}{0.700pt}}
\put(545,459){\rule[-0.350pt]{1.927pt}{0.700pt}}
\put(553,460){\rule[-0.350pt]{1.927pt}{0.700pt}}
\put(561,461){\rule[-0.350pt]{1.927pt}{0.700pt}}
\put(569,462){\rule[-0.350pt]{2.770pt}{0.700pt}}
\put(580,463){\rule[-0.350pt]{2.770pt}{0.700pt}}
\put(592,464){\rule[-0.350pt]{1.927pt}{0.700pt}}
\put(600,465){\rule[-0.350pt]{1.927pt}{0.700pt}}
\put(608,466){\rule[-0.350pt]{1.927pt}{0.700pt}}
\put(616,467){\rule[-0.350pt]{1.847pt}{0.700pt}}
\put(623,468){\rule[-0.350pt]{1.847pt}{0.700pt}}
\put(631,469){\rule[-0.350pt]{1.847pt}{0.700pt}}
\put(639,470){\rule[-0.350pt]{1.847pt}{0.700pt}}
\put(646,471){\rule[-0.350pt]{1.847pt}{0.700pt}}
\put(654,472){\rule[-0.350pt]{1.847pt}{0.700pt}}
\put(662,473){\rule[-0.350pt]{2.891pt}{0.700pt}}
\put(674,474){\rule[-0.350pt]{2.891pt}{0.700pt}}
\put(686,475){\rule[-0.350pt]{1.847pt}{0.700pt}}
\put(693,476){\rule[-0.350pt]{1.847pt}{0.700pt}}
\put(701,477){\rule[-0.350pt]{1.847pt}{0.700pt}}
\put(709,478){\rule[-0.350pt]{1.927pt}{0.700pt}}
\put(717,479){\rule[-0.350pt]{1.927pt}{0.700pt}}
\put(725,480){\rule[-0.350pt]{1.927pt}{0.700pt}}
\put(733,481){\rule[-0.350pt]{2.770pt}{0.700pt}}
\put(744,482){\rule[-0.350pt]{2.770pt}{0.700pt}}
\put(756,483){\rule[-0.350pt]{1.927pt}{0.700pt}}
\put(764,484){\rule[-0.350pt]{1.927pt}{0.700pt}}
\put(772,485){\rule[-0.350pt]{1.927pt}{0.700pt}}
\put(780,486){\rule[-0.350pt]{1.847pt}{0.700pt}}
\put(787,487){\rule[-0.350pt]{1.847pt}{0.700pt}}
\put(795,488){\rule[-0.350pt]{1.847pt}{0.700pt}}
\put(803,489){\rule[-0.350pt]{2.891pt}{0.700pt}}
\put(815,490){\rule[-0.350pt]{2.891pt}{0.700pt}}
\put(827,491){\rule[-0.350pt]{1.847pt}{0.700pt}}
\put(834,492){\rule[-0.350pt]{1.847pt}{0.700pt}}
\put(842,493){\rule[-0.350pt]{1.847pt}{0.700pt}}
\put(850,494){\rule[-0.350pt]{2.770pt}{0.700pt}}
\put(861,495){\rule[-0.350pt]{2.770pt}{0.700pt}}
\put(873,496){\rule[-0.350pt]{1.927pt}{0.700pt}}
\put(881,497){\rule[-0.350pt]{1.927pt}{0.700pt}}
\put(889,498){\rule[-0.350pt]{1.927pt}{0.700pt}}
\put(897,499){\rule[-0.350pt]{2.770pt}{0.700pt}}
\put(908,500){\rule[-0.350pt]{2.770pt}{0.700pt}}
\put(920,501){\rule[-0.350pt]{2.891pt}{0.700pt}}
\put(932,502){\rule[-0.350pt]{2.891pt}{0.700pt}}
\put(944,503){\rule[-0.350pt]{1.847pt}{0.700pt}}
\put(951,504){\rule[-0.350pt]{1.847pt}{0.700pt}}
\put(959,505){\rule[-0.350pt]{1.847pt}{0.700pt}}
\put(967,506){\rule[-0.350pt]{2.891pt}{0.700pt}}
\put(979,507){\rule[-0.350pt]{2.891pt}{0.700pt}}
\put(991,508){\rule[-0.350pt]{2.770pt}{0.700pt}}
\put(1002,509){\rule[-0.350pt]{2.770pt}{0.700pt}}
\put(1014,510){\rule[-0.350pt]{1.927pt}{0.700pt}}
\put(1022,511){\rule[-0.350pt]{1.927pt}{0.700pt}}
\put(1030,512){\rule[-0.350pt]{1.927pt}{0.700pt}}
\put(1038,513){\rule[-0.350pt]{2.770pt}{0.700pt}}
\put(1049,514){\rule[-0.350pt]{2.770pt}{0.700pt}}
\put(1061,515){\rule[-0.350pt]{2.770pt}{0.700pt}}
\put(1072,516){\rule[-0.350pt]{2.770pt}{0.700pt}}
\put(1084,517){\rule[-0.350pt]{2.891pt}{0.700pt}}
\put(1096,518){\rule[-0.350pt]{2.891pt}{0.700pt}}
\put(1108,519){\rule[-0.350pt]{1.847pt}{0.700pt}}
\put(1115,520){\rule[-0.350pt]{1.847pt}{0.700pt}}
\put(1123,521){\rule[-0.350pt]{1.847pt}{0.700pt}}
\put(1130,522){\rule[-0.350pt]{2.891pt}{0.700pt}}
\put(1143,523){\rule[-0.350pt]{2.891pt}{0.700pt}}
\put(1155,524){\rule[-0.350pt]{2.770pt}{0.700pt}}
\put(1166,525){\rule[-0.350pt]{2.770pt}{0.700pt}}
\put(1178,526){\rule[-0.350pt]{2.891pt}{0.700pt}}
\put(1190,527){\rule[-0.350pt]{2.891pt}{0.700pt}}
\put(1202,528){\rule[-0.350pt]{2.770pt}{0.700pt}}
\put(1213,529){\rule[-0.350pt]{2.770pt}{0.700pt}}
\put(1225,530){\rule[-0.350pt]{2.770pt}{0.700pt}}
\put(1236,531){\rule[-0.350pt]{2.770pt}{0.700pt}}
\put(1248,532){\rule[-0.350pt]{2.891pt}{0.700pt}}
\put(1260,533){\rule[-0.350pt]{2.891pt}{0.700pt}}
\put(1272,534){\rule[-0.350pt]{2.770pt}{0.700pt}}
\put(1283,535){\rule[-0.350pt]{2.770pt}{0.700pt}}
\put(1295,536){\rule[-0.350pt]{2.891pt}{0.700pt}}
\put(1307,537){\rule[-0.350pt]{2.891pt}{0.700pt}}
\put(1319,538){\rule[-0.350pt]{2.770pt}{0.700pt}}
\put(1330,539){\rule[-0.350pt]{2.770pt}{0.700pt}}
\put(1342,540){\rule[-0.350pt]{2.891pt}{0.700pt}}
\put(1354,541){\rule[-0.350pt]{2.891pt}{0.700pt}}
\put(1366,542){\rule[-0.350pt]{2.770pt}{0.700pt}}
\put(1377,543){\rule[-0.350pt]{2.770pt}{0.700pt}}
\put(1389,544){\rule[-0.350pt]{2.891pt}{0.700pt}}
\put(1401,545){\rule[-0.350pt]{2.891pt}{0.700pt}}
\put(1413,546){\rule[-0.350pt]{2.770pt}{0.700pt}}
\put(1424,547){\rule[-0.350pt]{2.770pt}{0.700pt}}
\sbox{\plotpoint}{\rule[-0.500pt]{1.000pt}{1.000pt}}%
\put(1306,632){\makebox(0,0)[r]{$m_\pi^2~SD$}}
\put(1328,632){\rule[-0.500pt]{15.899pt}{1.000pt}}
\put(264,487){\usebox{\plotpoint}}
\put(264,487){\usebox{\plotpoint}}
\put(265,486){\usebox{\plotpoint}}
\put(267,485){\usebox{\plotpoint}}
\put(268,484){\usebox{\plotpoint}}
\put(270,483){\usebox{\plotpoint}}
\put(272,482){\usebox{\plotpoint}}
\put(273,481){\usebox{\plotpoint}}
\put(275,480){\usebox{\plotpoint}}
\put(277,479){\usebox{\plotpoint}}
\put(278,478){\usebox{\plotpoint}}
\put(280,477){\usebox{\plotpoint}}
\put(282,476){\usebox{\plotpoint}}
\put(283,475){\usebox{\plotpoint}}
\put(285,474){\usebox{\plotpoint}}
\put(286,473){\usebox{\plotpoint}}
\put(288,472){\usebox{\plotpoint}}
\put(290,471){\usebox{\plotpoint}}
\put(292,470){\usebox{\plotpoint}}
\put(294,469){\usebox{\plotpoint}}
\put(296,468){\usebox{\plotpoint}}
\put(298,467){\usebox{\plotpoint}}
\put(299,466){\usebox{\plotpoint}}
\put(301,465){\usebox{\plotpoint}}
\put(303,464){\usebox{\plotpoint}}
\put(305,463){\usebox{\plotpoint}}
\put(307,462){\usebox{\plotpoint}}
\put(309,461){\usebox{\plotpoint}}
\put(311,460){\usebox{\plotpoint}}
\put(312,459){\usebox{\plotpoint}}
\put(314,458){\usebox{\plotpoint}}
\put(316,457){\usebox{\plotpoint}}
\put(318,456){\usebox{\plotpoint}}
\put(320,455){\usebox{\plotpoint}}
\put(322,454){\usebox{\plotpoint}}
\put(324,453){\usebox{\plotpoint}}
\put(326,452){\usebox{\plotpoint}}
\put(328,451){\usebox{\plotpoint}}
\put(330,450){\usebox{\plotpoint}}
\put(332,449){\usebox{\plotpoint}}
\put(333,448){\usebox{\plotpoint}}
\put(336,447){\usebox{\plotpoint}}
\put(338,446){\usebox{\plotpoint}}
\put(340,445){\usebox{\plotpoint}}
\put(342,444){\usebox{\plotpoint}}
\put(344,443){\usebox{\plotpoint}}
\put(347,442){\usebox{\plotpoint}}
\put(349,441){\usebox{\plotpoint}}
\put(351,440){\usebox{\plotpoint}}
\put(353,439){\usebox{\plotpoint}}
\put(355,438){\usebox{\plotpoint}}
\put(358,437){\usebox{\plotpoint}}
\put(360,436){\usebox{\plotpoint}}
\put(362,435){\usebox{\plotpoint}}
\put(364,434){\usebox{\plotpoint}}
\put(367,433){\usebox{\plotpoint}}
\put(369,432){\usebox{\plotpoint}}
\put(371,431){\usebox{\plotpoint}}
\put(374,430){\usebox{\plotpoint}}
\put(376,429){\usebox{\plotpoint}}
\put(378,428){\usebox{\plotpoint}}
\put(380,427){\usebox{\plotpoint}}
\put(383,426){\usebox{\plotpoint}}
\put(385,425){\usebox{\plotpoint}}
\put(388,424){\usebox{\plotpoint}}
\put(390,423){\usebox{\plotpoint}}
\put(392,422){\usebox{\plotpoint}}
\put(395,421){\usebox{\plotpoint}}
\put(397,420){\usebox{\plotpoint}}
\put(400,419){\usebox{\plotpoint}}
\put(402,418){\usebox{\plotpoint}}
\put(404,417){\usebox{\plotpoint}}
\put(407,416){\usebox{\plotpoint}}
\put(409,415){\usebox{\plotpoint}}
\put(411,414){\usebox{\plotpoint}}
\put(414,413){\usebox{\plotpoint}}
\put(416,412){\usebox{\plotpoint}}
\put(418,411){\usebox{\plotpoint}}
\put(421,410){\usebox{\plotpoint}}
\put(423,409){\usebox{\plotpoint}}
\put(425,408){\usebox{\plotpoint}}
\put(427,407){\usebox{\plotpoint}}
\put(430,406){\usebox{\plotpoint}}
\put(433,405){\usebox{\plotpoint}}
\put(435,404){\usebox{\plotpoint}}
\put(438,403){\usebox{\plotpoint}}
\put(441,402){\usebox{\plotpoint}}
\put(443,401){\usebox{\plotpoint}}
\put(446,400){\usebox{\plotpoint}}
\put(449,399){\usebox{\plotpoint}}
\put(451,398){\usebox{\plotpoint}}
\put(454,397){\usebox{\plotpoint}}
\put(457,396){\usebox{\plotpoint}}
\put(460,395){\usebox{\plotpoint}}
\put(463,394){\usebox{\plotpoint}}
\put(466,393){\usebox{\plotpoint}}
\put(469,392){\usebox{\plotpoint}}
\put(472,391){\usebox{\plotpoint}}
\put(475,390){\usebox{\plotpoint}}
\put(477,389){\usebox{\plotpoint}}
\put(480,388){\usebox{\plotpoint}}
\put(483,387){\usebox{\plotpoint}}
\put(486,386){\usebox{\plotpoint}}
\put(489,385){\usebox{\plotpoint}}
\put(492,384){\usebox{\plotpoint}}
\put(495,383){\usebox{\plotpoint}}
\put(498,382){\usebox{\plotpoint}}
\put(501,381){\usebox{\plotpoint}}
\put(504,380){\usebox{\plotpoint}}
\put(507,379){\usebox{\plotpoint}}
\put(510,378){\usebox{\plotpoint}}
\put(513,377){\usebox{\plotpoint}}
\put(516,376){\usebox{\plotpoint}}
\put(519,375){\usebox{\plotpoint}}
\put(522,374){\usebox{\plotpoint}}
\put(525,373){\usebox{\plotpoint}}
\put(528,372){\usebox{\plotpoint}}
\put(531,371){\usebox{\plotpoint}}
\put(535,370){\usebox{\plotpoint}}
\put(538,369){\usebox{\plotpoint}}
\put(541,368){\usebox{\plotpoint}}
\put(544,367){\usebox{\plotpoint}}
\put(548,366){\usebox{\plotpoint}}
\put(551,365){\usebox{\plotpoint}}
\put(555,364){\usebox{\plotpoint}}
\put(558,363){\usebox{\plotpoint}}
\put(562,362){\usebox{\plotpoint}}
\put(565,361){\usebox{\plotpoint}}
\put(569,360){\usebox{\plotpoint}}
\put(572,359){\usebox{\plotpoint}}
\put(575,358){\usebox{\plotpoint}}
\put(578,357){\usebox{\plotpoint}}
\put(582,356){\usebox{\plotpoint}}
\put(585,355){\usebox{\plotpoint}}
\put(588,354){\usebox{\plotpoint}}
\put(591,353){\usebox{\plotpoint}}
\put(596,352){\usebox{\plotpoint}}
\put(600,351){\usebox{\plotpoint}}
\put(604,350){\usebox{\plotpoint}}
\put(608,349){\usebox{\plotpoint}}
\put(612,348){\usebox{\plotpoint}}
\put(616,347){\usebox{\plotpoint}}
\put(619,346){\usebox{\plotpoint}}
\put(623,345){\usebox{\plotpoint}}
\put(627,344){\usebox{\plotpoint}}
\put(631,343){\usebox{\plotpoint}}
\put(635,342){\usebox{\plotpoint}}
\put(638,341){\usebox{\plotpoint}}
\put(642,340){\usebox{\plotpoint}}
\put(646,339){\usebox{\plotpoint}}
\put(650,338){\usebox{\plotpoint}}
\put(654,337){\usebox{\plotpoint}}
\put(658,336){\usebox{\plotpoint}}
\put(661,335){\rule[-0.500pt]{1.156pt}{1.000pt}}
\put(666,334){\rule[-0.500pt]{1.156pt}{1.000pt}}
\put(671,333){\rule[-0.500pt]{1.156pt}{1.000pt}}
\put(676,332){\rule[-0.500pt]{1.156pt}{1.000pt}}
\put(681,331){\rule[-0.500pt]{1.156pt}{1.000pt}}
\put(685,330){\rule[-0.500pt]{1.108pt}{1.000pt}}
\put(690,329){\rule[-0.500pt]{1.108pt}{1.000pt}}
\put(695,328){\rule[-0.500pt]{1.108pt}{1.000pt}}
\put(699,327){\rule[-0.500pt]{1.108pt}{1.000pt}}
\put(704,326){\rule[-0.500pt]{1.108pt}{1.000pt}}
\put(708,325){\rule[-0.500pt]{1.156pt}{1.000pt}}
\put(713,324){\rule[-0.500pt]{1.156pt}{1.000pt}}
\put(718,323){\rule[-0.500pt]{1.156pt}{1.000pt}}
\put(723,322){\rule[-0.500pt]{1.156pt}{1.000pt}}
\put(728,321){\rule[-0.500pt]{1.156pt}{1.000pt}}
\put(732,320){\rule[-0.500pt]{1.108pt}{1.000pt}}
\put(737,319){\rule[-0.500pt]{1.108pt}{1.000pt}}
\put(742,318){\rule[-0.500pt]{1.108pt}{1.000pt}}
\put(746,317){\rule[-0.500pt]{1.108pt}{1.000pt}}
\put(751,316){\rule[-0.500pt]{1.108pt}{1.000pt}}
\put(755,315){\rule[-0.500pt]{1.445pt}{1.000pt}}
\put(762,314){\rule[-0.500pt]{1.445pt}{1.000pt}}
\put(768,313){\rule[-0.500pt]{1.445pt}{1.000pt}}
\put(774,312){\rule[-0.500pt]{1.445pt}{1.000pt}}
\put(780,311){\rule[-0.500pt]{1.108pt}{1.000pt}}
\put(784,310){\rule[-0.500pt]{1.108pt}{1.000pt}}
\put(789,309){\rule[-0.500pt]{1.108pt}{1.000pt}}
\put(793,308){\rule[-0.500pt]{1.108pt}{1.000pt}}
\put(798,307){\rule[-0.500pt]{1.108pt}{1.000pt}}
\put(802,306){\rule[-0.500pt]{1.445pt}{1.000pt}}
\put(809,305){\rule[-0.500pt]{1.445pt}{1.000pt}}
\put(815,304){\rule[-0.500pt]{1.445pt}{1.000pt}}
\put(821,303){\rule[-0.500pt]{1.445pt}{1.000pt}}
\put(827,302){\rule[-0.500pt]{1.385pt}{1.000pt}}
\put(832,301){\rule[-0.500pt]{1.385pt}{1.000pt}}
\put(838,300){\rule[-0.500pt]{1.385pt}{1.000pt}}
\put(844,299){\rule[-0.500pt]{1.385pt}{1.000pt}}
\put(850,298){\rule[-0.500pt]{1.385pt}{1.000pt}}
\put(855,297){\rule[-0.500pt]{1.385pt}{1.000pt}}
\put(861,296){\rule[-0.500pt]{1.385pt}{1.000pt}}
\put(867,295){\rule[-0.500pt]{1.385pt}{1.000pt}}
\put(873,294){\rule[-0.500pt]{1.927pt}{1.000pt}}
\put(881,293){\rule[-0.500pt]{1.927pt}{1.000pt}}
\put(889,292){\rule[-0.500pt]{1.927pt}{1.000pt}}
\put(897,291){\rule[-0.500pt]{1.385pt}{1.000pt}}
\put(902,290){\rule[-0.500pt]{1.385pt}{1.000pt}}
\put(908,289){\rule[-0.500pt]{1.385pt}{1.000pt}}
\put(914,288){\rule[-0.500pt]{1.385pt}{1.000pt}}
\put(920,287){\rule[-0.500pt]{1.927pt}{1.000pt}}
\put(928,286){\rule[-0.500pt]{1.927pt}{1.000pt}}
\put(936,285){\rule[-0.500pt]{1.927pt}{1.000pt}}
\put(944,284){\rule[-0.500pt]{1.847pt}{1.000pt}}
\put(951,283){\rule[-0.500pt]{1.847pt}{1.000pt}}
\put(959,282){\rule[-0.500pt]{1.847pt}{1.000pt}}
\put(967,281){\rule[-0.500pt]{1.445pt}{1.000pt}}
\put(973,280){\rule[-0.500pt]{1.445pt}{1.000pt}}
\put(979,279){\rule[-0.500pt]{1.445pt}{1.000pt}}
\put(985,278){\rule[-0.500pt]{1.445pt}{1.000pt}}
\put(991,277){\rule[-0.500pt]{1.847pt}{1.000pt}}
\put(998,276){\rule[-0.500pt]{1.847pt}{1.000pt}}
\put(1006,275){\rule[-0.500pt]{1.847pt}{1.000pt}}
\put(1014,274){\rule[-0.500pt]{2.891pt}{1.000pt}}
\put(1026,273){\rule[-0.500pt]{2.891pt}{1.000pt}}
\put(1038,272){\rule[-0.500pt]{1.847pt}{1.000pt}}
\put(1045,271){\rule[-0.500pt]{1.847pt}{1.000pt}}
\put(1053,270){\rule[-0.500pt]{1.847pt}{1.000pt}}
\put(1060,269){\rule[-0.500pt]{1.847pt}{1.000pt}}
\put(1068,268){\rule[-0.500pt]{1.847pt}{1.000pt}}
\put(1076,267){\rule[-0.500pt]{1.847pt}{1.000pt}}
\put(1083,266){\rule[-0.500pt]{1.927pt}{1.000pt}}
\put(1092,265){\rule[-0.500pt]{1.927pt}{1.000pt}}
\put(1100,264){\rule[-0.500pt]{1.927pt}{1.000pt}}
\put(1108,263){\rule[-0.500pt]{2.770pt}{1.000pt}}
\put(1119,262){\rule[-0.500pt]{2.770pt}{1.000pt}}
\put(1131,261){\rule[-0.500pt]{2.891pt}{1.000pt}}
\put(1143,260){\rule[-0.500pt]{2.891pt}{1.000pt}}
\put(1155,259){\rule[-0.500pt]{1.847pt}{1.000pt}}
\put(1162,258){\rule[-0.500pt]{1.847pt}{1.000pt}}
\put(1170,257){\rule[-0.500pt]{1.847pt}{1.000pt}}
\put(1177,256){\rule[-0.500pt]{2.891pt}{1.000pt}}
\put(1190,255){\rule[-0.500pt]{2.891pt}{1.000pt}}
\put(1202,254){\rule[-0.500pt]{2.770pt}{1.000pt}}
\put(1213,253){\rule[-0.500pt]{2.770pt}{1.000pt}}
\put(1225,252){\rule[-0.500pt]{2.770pt}{1.000pt}}
\put(1236,251){\rule[-0.500pt]{2.770pt}{1.000pt}}
\put(1248,250){\rule[-0.500pt]{2.891pt}{1.000pt}}
\put(1260,249){\rule[-0.500pt]{2.891pt}{1.000pt}}
\put(1272,248){\rule[-0.500pt]{2.770pt}{1.000pt}}
\put(1283,247){\rule[-0.500pt]{2.770pt}{1.000pt}}
\put(1295,246){\rule[-0.500pt]{2.891pt}{1.000pt}}
\put(1307,245){\rule[-0.500pt]{2.891pt}{1.000pt}}
\put(1319,244){\rule[-0.500pt]{2.770pt}{1.000pt}}
\put(1330,243){\rule[-0.500pt]{2.770pt}{1.000pt}}
\put(1342,242){\rule[-0.500pt]{2.891pt}{1.000pt}}
\put(1354,241){\rule[-0.500pt]{2.891pt}{1.000pt}}
\put(1366,240){\rule[-0.500pt]{2.770pt}{1.000pt}}
\put(1377,239){\rule[-0.500pt]{2.770pt}{1.000pt}}
\put(1389,238){\rule[-0.500pt]{5.782pt}{1.000pt}}
\put(1413,237){\rule[-0.500pt]{2.770pt}{1.000pt}}
\put(1424,236){\rule[-0.500pt]{2.770pt}{1.000pt}}
\sbox{\plotpoint}{\rule[-0.250pt]{0.500pt}{0.500pt}}%
\put(1306,587){\makebox(0,0)[r]{$\Delta D~SD$}}
\put(1328,587){\usebox{\plotpoint}}
\put(1348,587){\usebox{\plotpoint}}
\put(1369,587){\usebox{\plotpoint}}
\put(1390,587){\usebox{\plotpoint}}
\put(1394,587){\usebox{\plotpoint}}
\put(264,412){\usebox{\plotpoint}}
\put(264,412){\usebox{\plotpoint}}
\put(282,403){\usebox{\plotpoint}}
\put(301,394){\usebox{\plotpoint}}
\put(321,387){\usebox{\plotpoint}}
\put(340,379){\usebox{\plotpoint}}
\put(359,372){\usebox{\plotpoint}}
\put(379,365){\usebox{\plotpoint}}
\put(399,358){\usebox{\plotpoint}}
\put(418,352){\usebox{\plotpoint}}
\put(438,345){\usebox{\plotpoint}}
\put(458,340){\usebox{\plotpoint}}
\put(478,334){\usebox{\plotpoint}}
\put(498,328){\usebox{\plotpoint}}
\put(518,323){\usebox{\plotpoint}}
\put(538,319){\usebox{\plotpoint}}
\put(558,314){\usebox{\plotpoint}}
\put(579,309){\usebox{\plotpoint}}
\put(599,305){\usebox{\plotpoint}}
\put(619,301){\usebox{\plotpoint}}
\put(640,297){\usebox{\plotpoint}}
\put(660,293){\usebox{\plotpoint}}
\put(681,289){\usebox{\plotpoint}}
\put(701,286){\usebox{\plotpoint}}
\put(721,282){\usebox{\plotpoint}}
\put(742,279){\usebox{\plotpoint}}
\put(763,276){\usebox{\plotpoint}}
\put(783,273){\usebox{\plotpoint}}
\put(804,270){\usebox{\plotpoint}}
\put(824,268){\usebox{\plotpoint}}
\put(845,265){\usebox{\plotpoint}}
\put(865,262){\usebox{\plotpoint}}
\put(886,260){\usebox{\plotpoint}}
\put(907,257){\usebox{\plotpoint}}
\put(927,255){\usebox{\plotpoint}}
\put(948,253){\usebox{\plotpoint}}
\put(968,250){\usebox{\plotpoint}}
\put(989,249){\usebox{\plotpoint}}
\put(1010,246){\usebox{\plotpoint}}
\put(1030,244){\usebox{\plotpoint}}
\put(1051,242){\usebox{\plotpoint}}
\put(1072,241){\usebox{\plotpoint}}
\put(1092,239){\usebox{\plotpoint}}
\put(1113,237){\usebox{\plotpoint}}
\put(1134,235){\usebox{\plotpoint}}
\put(1154,234){\usebox{\plotpoint}}
\put(1175,232){\usebox{\plotpoint}}
\put(1196,231){\usebox{\plotpoint}}
\put(1217,229){\usebox{\plotpoint}}
\put(1237,227){\usebox{\plotpoint}}
\put(1258,226){\usebox{\plotpoint}}
\put(1279,225){\usebox{\plotpoint}}
\put(1299,223){\usebox{\plotpoint}}
\put(1320,222){\usebox{\plotpoint}}
\put(1341,221){\usebox{\plotpoint}}
\put(1361,220){\usebox{\plotpoint}}
\put(1382,219){\usebox{\plotpoint}}
\put(1403,218){\usebox{\plotpoint}}
\put(1424,217){\usebox{\plotpoint}}
\put(1436,216){\usebox{\plotpoint}}
\end{picture}

\caption[Long and short distance contributions to the mass-squared differences
as a function of the matching scale~$\Lambda$.
Plotted are $\dem{\pi}$ and the correction needed for Dashen's theorem.]
{Long and short distance contributions to the mass-squared differences
as a function of the matching scale~$\Lambda$.
Plotted are $\dem{\pi}$ and the correction needed for Dashen's theorem,
$\Delta D =
\left(m_{K^+}^2 - m_{K^0}^2 -m_{\pi^+}^2 + m_{\pi^0}^2\right)_{em}$.}
\label{fig3}
\end{figure}
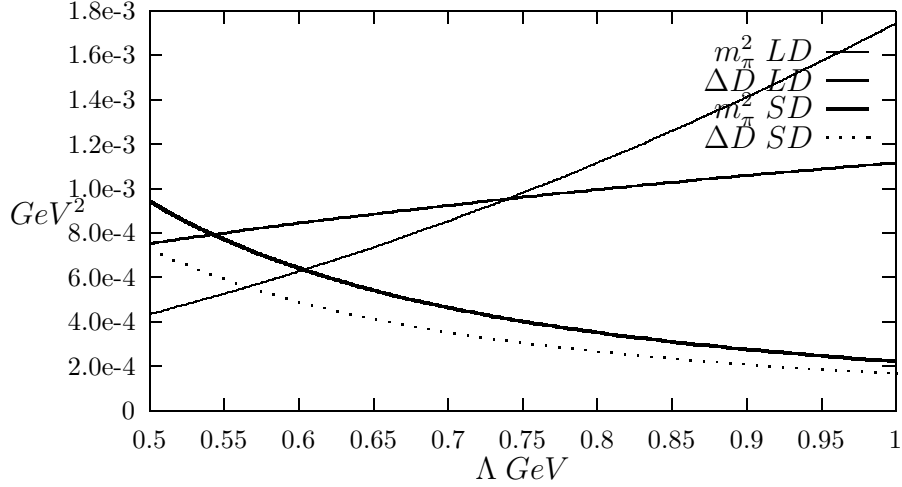
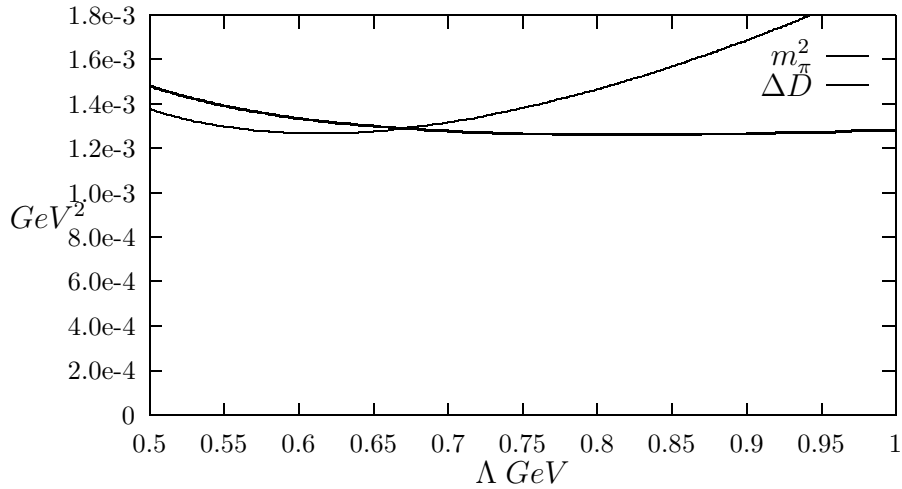
\begin{figure}
\setlength{\unitlength}{0.240900pt}
\ifx\plotpoint\undefined\newsavebox{\plotpoint}\fi
\sbox{\plotpoint}{\rule[-0.175pt]{0.350pt}{0.350pt}}%
\begin{picture}(1500,900)(0,0)
\tenrm
\sbox{\plotpoint}{\rule[-0.175pt]{0.350pt}{0.350pt}}%
\put(264,158){\rule[-0.175pt]{282.335pt}{0.350pt}}
\put(264,158){\rule[-0.175pt]{4.818pt}{0.350pt}}
\put(242,158){\makebox(0,0)[r]{0}}
\put(1416,158){\rule[-0.175pt]{4.818pt}{0.350pt}}
\put(264,228){\rule[-0.175pt]{4.818pt}{0.350pt}}
\put(242,228){\makebox(0,0)[r]{2.0e-4}}
\put(1416,228){\rule[-0.175pt]{4.818pt}{0.350pt}}
\put(264,298){\rule[-0.175pt]{4.818pt}{0.350pt}}
\put(242,298){\makebox(0,0)[r]{4.0e-4}}
\put(1416,298){\rule[-0.175pt]{4.818pt}{0.350pt}}
\put(264,368){\rule[-0.175pt]{4.818pt}{0.350pt}}
\put(242,368){\makebox(0,0)[r]{6.0e-4}}
\put(1416,368){\rule[-0.175pt]{4.818pt}{0.350pt}}
\put(264,438){\rule[-0.175pt]{4.818pt}{0.350pt}}
\put(242,438){\makebox(0,0)[r]{8.0e-4}}
\put(1416,438){\rule[-0.175pt]{4.818pt}{0.350pt}}
\put(264,507){\rule[-0.175pt]{4.818pt}{0.350pt}}
\put(242,507){\makebox(0,0)[r]{1.0e-3}}
\put(1416,507){\rule[-0.175pt]{4.818pt}{0.350pt}}
\put(264,577){\rule[-0.175pt]{4.818pt}{0.350pt}}
\put(242,577){\makebox(0,0)[r]{1.2e-3}}
\put(1416,577){\rule[-0.175pt]{4.818pt}{0.350pt}}
\put(264,647){\rule[-0.175pt]{4.818pt}{0.350pt}}
\put(242,647){\makebox(0,0)[r]{1.4e-3}}
\put(1416,647){\rule[-0.175pt]{4.818pt}{0.350pt}}
\put(264,717){\rule[-0.175pt]{4.818pt}{0.350pt}}
\put(242,717){\makebox(0,0)[r]{1.6e-3}}
\put(1416,717){\rule[-0.175pt]{4.818pt}{0.350pt}}
\put(264,787){\rule[-0.175pt]{4.818pt}{0.350pt}}
\put(242,787){\makebox(0,0)[r]{1.8e-3}}
\put(1416,787){\rule[-0.175pt]{4.818pt}{0.350pt}}
\put(264,158){\rule[-0.175pt]{0.350pt}{4.818pt}}
\put(264,113){\makebox(0,0){0.5}}
\put(264,767){\rule[-0.175pt]{0.350pt}{4.818pt}}
\put(381,158){\rule[-0.175pt]{0.350pt}{4.818pt}}
\put(381,113){\makebox(0,0){0.55}}
\put(381,767){\rule[-0.175pt]{0.350pt}{4.818pt}}
\put(498,158){\rule[-0.175pt]{0.350pt}{4.818pt}}
\put(498,113){\makebox(0,0){0.6}}
\put(498,767){\rule[-0.175pt]{0.350pt}{4.818pt}}
\put(616,158){\rule[-0.175pt]{0.350pt}{4.818pt}}
\put(616,113){\makebox(0,0){0.65}}
\put(616,767){\rule[-0.175pt]{0.350pt}{4.818pt}}
\put(733,158){\rule[-0.175pt]{0.350pt}{4.818pt}}
\put(733,113){\makebox(0,0){0.7}}
\put(733,767){\rule[-0.175pt]{0.350pt}{4.818pt}}
\put(850,158){\rule[-0.175pt]{0.350pt}{4.818pt}}
\put(850,113){\makebox(0,0){0.75}}
\put(850,767){\rule[-0.175pt]{0.350pt}{4.818pt}}
\put(967,158){\rule[-0.175pt]{0.350pt}{4.818pt}}
\put(967,113){\makebox(0,0){0.8}}
\put(967,767){\rule[-0.175pt]{0.350pt}{4.818pt}}
\put(1084,158){\rule[-0.175pt]{0.350pt}{4.818pt}}
\put(1084,113){\makebox(0,0){0.85}}
\put(1084,767){\rule[-0.175pt]{0.350pt}{4.818pt}}
\put(1202,158){\rule[-0.175pt]{0.350pt}{4.818pt}}
\put(1202,113){\makebox(0,0){0.9}}
\put(1202,767){\rule[-0.175pt]{0.350pt}{4.818pt}}
\put(1319,158){\rule[-0.175pt]{0.350pt}{4.818pt}}
\put(1319,113){\makebox(0,0){0.95}}
\put(1319,767){\rule[-0.175pt]{0.350pt}{4.818pt}}
\put(1436,158){\rule[-0.175pt]{0.350pt}{4.818pt}}
\put(1436,113){\makebox(0,0){1}}
\put(1436,767){\rule[-0.175pt]{0.350pt}{4.818pt}}
\put(264,158){\rule[-0.175pt]{282.335pt}{0.350pt}}
\put(1436,158){\rule[-0.175pt]{0.350pt}{151.526pt}}
\put(264,787){\rule[-0.175pt]{282.335pt}{0.350pt}}
\put(45,472){\makebox(0,0)[l]{\shortstack{$GeV^2$}}}
\put(850,68){\makebox(0,0){$\Lambda~GeV$}}
\put(264,158){\rule[-0.175pt]{0.350pt}{151.526pt}}
\put(1306,722){\makebox(0,0)[r]{$m_\pi^2$}}
\put(1328,722){\rule[-0.175pt]{15.899pt}{0.350pt}}
\put(264,639){\usebox{\plotpoint}}
\put(264,639){\rule[-0.175pt]{0.693pt}{0.350pt}}
\put(266,638){\rule[-0.175pt]{0.693pt}{0.350pt}}
\put(269,637){\rule[-0.175pt]{0.693pt}{0.350pt}}
\put(272,636){\rule[-0.175pt]{0.693pt}{0.350pt}}
\put(275,635){\rule[-0.175pt]{0.693pt}{0.350pt}}
\put(278,634){\rule[-0.175pt]{0.693pt}{0.350pt}}
\put(281,633){\rule[-0.175pt]{0.693pt}{0.350pt}}
\put(284,632){\rule[-0.175pt]{0.693pt}{0.350pt}}
\put(287,631){\rule[-0.175pt]{0.964pt}{0.350pt}}
\put(291,630){\rule[-0.175pt]{0.964pt}{0.350pt}}
\put(295,629){\rule[-0.175pt]{0.964pt}{0.350pt}}
\put(299,628){\rule[-0.175pt]{0.964pt}{0.350pt}}
\put(303,627){\rule[-0.175pt]{0.964pt}{0.350pt}}
\put(307,626){\rule[-0.175pt]{0.964pt}{0.350pt}}
\put(311,625){\rule[-0.175pt]{0.923pt}{0.350pt}}
\put(314,624){\rule[-0.175pt]{0.923pt}{0.350pt}}
\put(318,623){\rule[-0.175pt]{0.923pt}{0.350pt}}
\put(322,622){\rule[-0.175pt]{0.923pt}{0.350pt}}
\put(326,621){\rule[-0.175pt]{0.923pt}{0.350pt}}
\put(330,620){\rule[-0.175pt]{0.923pt}{0.350pt}}
\put(334,619){\rule[-0.175pt]{1.445pt}{0.350pt}}
\put(340,618){\rule[-0.175pt]{1.445pt}{0.350pt}}
\put(346,617){\rule[-0.175pt]{1.445pt}{0.350pt}}
\put(352,616){\rule[-0.175pt]{1.445pt}{0.350pt}}
\put(358,615){\rule[-0.175pt]{1.385pt}{0.350pt}}
\put(363,614){\rule[-0.175pt]{1.385pt}{0.350pt}}
\put(369,613){\rule[-0.175pt]{1.385pt}{0.350pt}}
\put(375,612){\rule[-0.175pt]{1.385pt}{0.350pt}}
\put(381,611){\rule[-0.175pt]{1.927pt}{0.350pt}}
\put(389,610){\rule[-0.175pt]{1.927pt}{0.350pt}}
\put(397,609){\rule[-0.175pt]{1.927pt}{0.350pt}}
\put(405,608){\rule[-0.175pt]{1.847pt}{0.350pt}}
\put(412,607){\rule[-0.175pt]{1.847pt}{0.350pt}}
\put(420,606){\rule[-0.175pt]{1.847pt}{0.350pt}}
\put(427,605){\rule[-0.175pt]{2.891pt}{0.350pt}}
\put(440,604){\rule[-0.175pt]{2.891pt}{0.350pt}}
\put(452,603){\rule[-0.175pt]{5.541pt}{0.350pt}}
\put(475,602){\rule[-0.175pt]{5.541pt}{0.350pt}}
\put(498,601){\rule[-0.175pt]{17.104pt}{0.350pt}}
\put(569,602){\rule[-0.175pt]{5.541pt}{0.350pt}}
\put(592,603){\rule[-0.175pt]{5.782pt}{0.350pt}}
\put(616,604){\rule[-0.175pt]{2.770pt}{0.350pt}}
\put(627,605){\rule[-0.175pt]{2.770pt}{0.350pt}}
\put(639,606){\rule[-0.175pt]{1.847pt}{0.350pt}}
\put(646,607){\rule[-0.175pt]{1.847pt}{0.350pt}}
\put(654,608){\rule[-0.175pt]{1.847pt}{0.350pt}}
\put(662,609){\rule[-0.175pt]{1.927pt}{0.350pt}}
\put(670,610){\rule[-0.175pt]{1.927pt}{0.350pt}}
\put(678,611){\rule[-0.175pt]{1.927pt}{0.350pt}}
\put(686,612){\rule[-0.175pt]{1.847pt}{0.350pt}}
\put(693,613){\rule[-0.175pt]{1.847pt}{0.350pt}}
\put(701,614){\rule[-0.175pt]{1.847pt}{0.350pt}}
\put(709,615){\rule[-0.175pt]{1.927pt}{0.350pt}}
\put(717,616){\rule[-0.175pt]{1.927pt}{0.350pt}}
\put(725,617){\rule[-0.175pt]{1.927pt}{0.350pt}}
\put(733,618){\rule[-0.175pt]{1.385pt}{0.350pt}}
\put(738,619){\rule[-0.175pt]{1.385pt}{0.350pt}}
\put(744,620){\rule[-0.175pt]{1.385pt}{0.350pt}}
\put(750,621){\rule[-0.175pt]{1.385pt}{0.350pt}}
\put(756,622){\rule[-0.175pt]{1.445pt}{0.350pt}}
\put(762,623){\rule[-0.175pt]{1.445pt}{0.350pt}}
\put(768,624){\rule[-0.175pt]{1.445pt}{0.350pt}}
\put(774,625){\rule[-0.175pt]{1.445pt}{0.350pt}}
\put(780,626){\rule[-0.175pt]{1.108pt}{0.350pt}}
\put(784,627){\rule[-0.175pt]{1.108pt}{0.350pt}}
\put(789,628){\rule[-0.175pt]{1.108pt}{0.350pt}}
\put(793,629){\rule[-0.175pt]{1.108pt}{0.350pt}}
\put(798,630){\rule[-0.175pt]{1.108pt}{0.350pt}}
\put(802,631){\rule[-0.175pt]{1.156pt}{0.350pt}}
\put(807,632){\rule[-0.175pt]{1.156pt}{0.350pt}}
\put(812,633){\rule[-0.175pt]{1.156pt}{0.350pt}}
\put(817,634){\rule[-0.175pt]{1.156pt}{0.350pt}}
\put(822,635){\rule[-0.175pt]{1.156pt}{0.350pt}}
\put(826,636){\rule[-0.175pt]{1.108pt}{0.350pt}}
\put(831,637){\rule[-0.175pt]{1.108pt}{0.350pt}}
\put(836,638){\rule[-0.175pt]{1.108pt}{0.350pt}}
\put(840,639){\rule[-0.175pt]{1.108pt}{0.350pt}}
\put(845,640){\rule[-0.175pt]{1.108pt}{0.350pt}}
\put(849,641){\rule[-0.175pt]{1.108pt}{0.350pt}}
\put(854,642){\rule[-0.175pt]{1.108pt}{0.350pt}}
\put(859,643){\rule[-0.175pt]{1.108pt}{0.350pt}}
\put(863,644){\rule[-0.175pt]{1.108pt}{0.350pt}}
\put(868,645){\rule[-0.175pt]{1.108pt}{0.350pt}}
\put(872,646){\rule[-0.175pt]{0.964pt}{0.350pt}}
\put(877,647){\rule[-0.175pt]{0.964pt}{0.350pt}}
\put(881,648){\rule[-0.175pt]{0.964pt}{0.350pt}}
\put(885,649){\rule[-0.175pt]{0.964pt}{0.350pt}}
\put(889,650){\rule[-0.175pt]{0.964pt}{0.350pt}}
\put(893,651){\rule[-0.175pt]{0.964pt}{0.350pt}}
\put(897,652){\rule[-0.175pt]{0.923pt}{0.350pt}}
\put(900,653){\rule[-0.175pt]{0.923pt}{0.350pt}}
\put(904,654){\rule[-0.175pt]{0.923pt}{0.350pt}}
\put(908,655){\rule[-0.175pt]{0.923pt}{0.350pt}}
\put(912,656){\rule[-0.175pt]{0.923pt}{0.350pt}}
\put(916,657){\rule[-0.175pt]{0.923pt}{0.350pt}}
\put(919,658){\rule[-0.175pt]{0.964pt}{0.350pt}}
\put(924,659){\rule[-0.175pt]{0.964pt}{0.350pt}}
\put(928,660){\rule[-0.175pt]{0.964pt}{0.350pt}}
\put(932,661){\rule[-0.175pt]{0.964pt}{0.350pt}}
\put(936,662){\rule[-0.175pt]{0.964pt}{0.350pt}}
\put(940,663){\rule[-0.175pt]{0.964pt}{0.350pt}}
\put(944,664){\rule[-0.175pt]{0.923pt}{0.350pt}}
\put(947,665){\rule[-0.175pt]{0.923pt}{0.350pt}}
\put(951,666){\rule[-0.175pt]{0.923pt}{0.350pt}}
\put(955,667){\rule[-0.175pt]{0.923pt}{0.350pt}}
\put(959,668){\rule[-0.175pt]{0.923pt}{0.350pt}}
\put(963,669){\rule[-0.175pt]{0.923pt}{0.350pt}}
\put(966,670){\rule[-0.175pt]{0.826pt}{0.350pt}}
\put(970,671){\rule[-0.175pt]{0.826pt}{0.350pt}}
\put(973,672){\rule[-0.175pt]{0.826pt}{0.350pt}}
\put(977,673){\rule[-0.175pt]{0.826pt}{0.350pt}}
\put(980,674){\rule[-0.175pt]{0.826pt}{0.350pt}}
\put(984,675){\rule[-0.175pt]{0.826pt}{0.350pt}}
\put(987,676){\rule[-0.175pt]{0.826pt}{0.350pt}}
\put(991,677){\rule[-0.175pt]{0.792pt}{0.350pt}}
\put(994,678){\rule[-0.175pt]{0.792pt}{0.350pt}}
\put(997,679){\rule[-0.175pt]{0.792pt}{0.350pt}}
\put(1000,680){\rule[-0.175pt]{0.792pt}{0.350pt}}
\put(1004,681){\rule[-0.175pt]{0.792pt}{0.350pt}}
\put(1007,682){\rule[-0.175pt]{0.792pt}{0.350pt}}
\put(1010,683){\rule[-0.175pt]{0.792pt}{0.350pt}}
\put(1013,684){\rule[-0.175pt]{0.826pt}{0.350pt}}
\put(1017,685){\rule[-0.175pt]{0.826pt}{0.350pt}}
\put(1020,686){\rule[-0.175pt]{0.826pt}{0.350pt}}
\put(1024,687){\rule[-0.175pt]{0.826pt}{0.350pt}}
\put(1027,688){\rule[-0.175pt]{0.826pt}{0.350pt}}
\put(1031,689){\rule[-0.175pt]{0.826pt}{0.350pt}}
\put(1034,690){\rule[-0.175pt]{0.826pt}{0.350pt}}
\put(1038,691){\rule[-0.175pt]{0.792pt}{0.350pt}}
\put(1041,692){\rule[-0.175pt]{0.792pt}{0.350pt}}
\put(1044,693){\rule[-0.175pt]{0.792pt}{0.350pt}}
\put(1047,694){\rule[-0.175pt]{0.792pt}{0.350pt}}
\put(1051,695){\rule[-0.175pt]{0.792pt}{0.350pt}}
\put(1054,696){\rule[-0.175pt]{0.792pt}{0.350pt}}
\put(1057,697){\rule[-0.175pt]{0.791pt}{0.350pt}}
\put(1061,698){\rule[-0.175pt]{0.693pt}{0.350pt}}
\put(1063,699){\rule[-0.175pt]{0.693pt}{0.350pt}}
\put(1066,700){\rule[-0.175pt]{0.693pt}{0.350pt}}
\put(1069,701){\rule[-0.175pt]{0.693pt}{0.350pt}}
\put(1072,702){\rule[-0.175pt]{0.693pt}{0.350pt}}
\put(1075,703){\rule[-0.175pt]{0.693pt}{0.350pt}}
\put(1078,704){\rule[-0.175pt]{0.693pt}{0.350pt}}
\put(1081,705){\rule[-0.175pt]{0.693pt}{0.350pt}}
\put(1084,706){\rule[-0.175pt]{0.723pt}{0.350pt}}
\put(1087,707){\rule[-0.175pt]{0.723pt}{0.350pt}}
\put(1090,708){\rule[-0.175pt]{0.723pt}{0.350pt}}
\put(1093,709){\rule[-0.175pt]{0.723pt}{0.350pt}}
\put(1096,710){\rule[-0.175pt]{0.723pt}{0.350pt}}
\put(1099,711){\rule[-0.175pt]{0.723pt}{0.350pt}}
\put(1102,712){\rule[-0.175pt]{0.723pt}{0.350pt}}
\put(1105,713){\rule[-0.175pt]{0.723pt}{0.350pt}}
\put(1108,714){\rule[-0.175pt]{0.693pt}{0.350pt}}
\put(1110,715){\rule[-0.175pt]{0.693pt}{0.350pt}}
\put(1113,716){\rule[-0.175pt]{0.693pt}{0.350pt}}
\put(1116,717){\rule[-0.175pt]{0.693pt}{0.350pt}}
\put(1119,718){\rule[-0.175pt]{0.693pt}{0.350pt}}
\put(1122,719){\rule[-0.175pt]{0.693pt}{0.350pt}}
\put(1125,720){\rule[-0.175pt]{0.693pt}{0.350pt}}
\put(1128,721){\rule[-0.175pt]{0.693pt}{0.350pt}}
\put(1131,722){\rule[-0.175pt]{0.723pt}{0.350pt}}
\put(1134,723){\rule[-0.175pt]{0.723pt}{0.350pt}}
\put(1137,724){\rule[-0.175pt]{0.723pt}{0.350pt}}
\put(1140,725){\rule[-0.175pt]{0.723pt}{0.350pt}}
\put(1143,726){\rule[-0.175pt]{0.723pt}{0.350pt}}
\put(1146,727){\rule[-0.175pt]{0.723pt}{0.350pt}}
\put(1149,728){\rule[-0.175pt]{0.723pt}{0.350pt}}
\put(1152,729){\rule[-0.175pt]{0.723pt}{0.350pt}}
\put(1155,730){\rule[-0.175pt]{0.616pt}{0.350pt}}
\put(1157,731){\rule[-0.175pt]{0.616pt}{0.350pt}}
\put(1160,732){\rule[-0.175pt]{0.616pt}{0.350pt}}
\put(1162,733){\rule[-0.175pt]{0.616pt}{0.350pt}}
\put(1165,734){\rule[-0.175pt]{0.616pt}{0.350pt}}
\put(1167,735){\rule[-0.175pt]{0.616pt}{0.350pt}}
\put(1170,736){\rule[-0.175pt]{0.616pt}{0.350pt}}
\put(1172,737){\rule[-0.175pt]{0.616pt}{0.350pt}}
\put(1175,738){\rule[-0.175pt]{0.616pt}{0.350pt}}
\put(1177,739){\rule[-0.175pt]{0.723pt}{0.350pt}}
\put(1181,740){\rule[-0.175pt]{0.723pt}{0.350pt}}
\put(1184,741){\rule[-0.175pt]{0.723pt}{0.350pt}}
\put(1187,742){\rule[-0.175pt]{0.723pt}{0.350pt}}
\put(1190,743){\rule[-0.175pt]{0.723pt}{0.350pt}}
\put(1193,744){\rule[-0.175pt]{0.723pt}{0.350pt}}
\put(1196,745){\rule[-0.175pt]{0.723pt}{0.350pt}}
\put(1199,746){\rule[-0.175pt]{0.723pt}{0.350pt}}
\put(1202,747){\rule[-0.175pt]{0.616pt}{0.350pt}}
\put(1204,748){\rule[-0.175pt]{0.616pt}{0.350pt}}
\put(1207,749){\rule[-0.175pt]{0.616pt}{0.350pt}}
\put(1209,750){\rule[-0.175pt]{0.616pt}{0.350pt}}
\put(1212,751){\rule[-0.175pt]{0.616pt}{0.350pt}}
\put(1214,752){\rule[-0.175pt]{0.616pt}{0.350pt}}
\put(1217,753){\rule[-0.175pt]{0.616pt}{0.350pt}}
\put(1219,754){\rule[-0.175pt]{0.616pt}{0.350pt}}
\put(1222,755){\rule[-0.175pt]{0.616pt}{0.350pt}}
\put(1224,756){\rule[-0.175pt]{0.616pt}{0.350pt}}
\put(1227,757){\rule[-0.175pt]{0.616pt}{0.350pt}}
\put(1230,758){\rule[-0.175pt]{0.616pt}{0.350pt}}
\put(1232,759){\rule[-0.175pt]{0.616pt}{0.350pt}}
\put(1235,760){\rule[-0.175pt]{0.616pt}{0.350pt}}
\put(1237,761){\rule[-0.175pt]{0.616pt}{0.350pt}}
\put(1240,762){\rule[-0.175pt]{0.616pt}{0.350pt}}
\put(1242,763){\rule[-0.175pt]{0.616pt}{0.350pt}}
\put(1245,764){\rule[-0.175pt]{0.616pt}{0.350pt}}
\put(1247,765){\rule[-0.175pt]{0.642pt}{0.350pt}}
\put(1250,766){\rule[-0.175pt]{0.642pt}{0.350pt}}
\put(1253,767){\rule[-0.175pt]{0.642pt}{0.350pt}}
\put(1255,768){\rule[-0.175pt]{0.642pt}{0.350pt}}
\put(1258,769){\rule[-0.175pt]{0.642pt}{0.350pt}}
\put(1261,770){\rule[-0.175pt]{0.642pt}{0.350pt}}
\put(1263,771){\rule[-0.175pt]{0.642pt}{0.350pt}}
\put(1266,772){\rule[-0.175pt]{0.642pt}{0.350pt}}
\put(1269,773){\rule[-0.175pt]{0.642pt}{0.350pt}}
\put(1271,774){\rule[-0.175pt]{0.554pt}{0.350pt}}
\put(1274,775){\rule[-0.175pt]{0.554pt}{0.350pt}}
\put(1276,776){\rule[-0.175pt]{0.554pt}{0.350pt}}
\put(1278,777){\rule[-0.175pt]{0.554pt}{0.350pt}}
\put(1281,778){\rule[-0.175pt]{0.554pt}{0.350pt}}
\put(1283,779){\rule[-0.175pt]{0.554pt}{0.350pt}}
\put(1285,780){\rule[-0.175pt]{0.554pt}{0.350pt}}
\put(1288,781){\rule[-0.175pt]{0.554pt}{0.350pt}}
\put(1290,782){\rule[-0.175pt]{0.554pt}{0.350pt}}
\put(1292,783){\rule[-0.175pt]{0.554pt}{0.350pt}}
\put(1295,784){\rule[-0.175pt]{0.642pt}{0.350pt}}
\put(1297,785){\rule[-0.175pt]{0.642pt}{0.350pt}}
\put(1300,786){\rule[-0.175pt]{0.642pt}{0.350pt}}
\put(1302,787){\usebox{\plotpoint}}
\sbox{\plotpoint}{\rule[-0.350pt]{0.700pt}{0.700pt}}%
\put(1306,677){\makebox(0,0)[r]{$\Delta D$}}
\put(1328,677){\rule[-0.350pt]{15.899pt}{0.700pt}}
\put(264,675){\usebox{\plotpoint}}
\put(264,675){\rule[-0.350pt]{0.792pt}{0.700pt}}
\put(267,674){\rule[-0.350pt]{0.792pt}{0.700pt}}
\put(270,673){\rule[-0.350pt]{0.792pt}{0.700pt}}
\put(273,672){\rule[-0.350pt]{0.792pt}{0.700pt}}
\put(277,671){\rule[-0.350pt]{0.792pt}{0.700pt}}
\put(280,670){\rule[-0.350pt]{0.792pt}{0.700pt}}
\put(283,669){\rule[-0.350pt]{0.792pt}{0.700pt}}
\put(286,668){\rule[-0.350pt]{0.826pt}{0.700pt}}
\put(290,667){\rule[-0.350pt]{0.826pt}{0.700pt}}
\put(293,666){\rule[-0.350pt]{0.826pt}{0.700pt}}
\put(297,665){\rule[-0.350pt]{0.826pt}{0.700pt}}
\put(300,664){\rule[-0.350pt]{0.826pt}{0.700pt}}
\put(304,663){\rule[-0.350pt]{0.826pt}{0.700pt}}
\put(307,662){\rule[-0.350pt]{0.826pt}{0.700pt}}
\put(310,661){\rule[-0.350pt]{0.923pt}{0.700pt}}
\put(314,660){\rule[-0.350pt]{0.923pt}{0.700pt}}
\put(318,659){\rule[-0.350pt]{0.923pt}{0.700pt}}
\put(322,658){\rule[-0.350pt]{0.923pt}{0.700pt}}
\put(326,657){\rule[-0.350pt]{0.923pt}{0.700pt}}
\put(330,656){\rule[-0.350pt]{0.923pt}{0.700pt}}
\put(334,655){\rule[-0.350pt]{0.964pt}{0.700pt}}
\put(338,654){\rule[-0.350pt]{0.964pt}{0.700pt}}
\put(342,653){\rule[-0.350pt]{0.964pt}{0.700pt}}
\put(346,652){\rule[-0.350pt]{0.964pt}{0.700pt}}
\put(350,651){\rule[-0.350pt]{0.964pt}{0.700pt}}
\put(354,650){\rule[-0.350pt]{0.964pt}{0.700pt}}
\put(358,649){\rule[-0.350pt]{1.108pt}{0.700pt}}
\put(362,648){\rule[-0.350pt]{1.108pt}{0.700pt}}
\put(367,647){\rule[-0.350pt]{1.108pt}{0.700pt}}
\put(371,646){\rule[-0.350pt]{1.108pt}{0.700pt}}
\put(376,645){\rule[-0.350pt]{1.108pt}{0.700pt}}
\put(381,644){\rule[-0.350pt]{1.156pt}{0.700pt}}
\put(385,643){\rule[-0.350pt]{1.156pt}{0.700pt}}
\put(390,642){\rule[-0.350pt]{1.156pt}{0.700pt}}
\put(395,641){\rule[-0.350pt]{1.156pt}{0.700pt}}
\put(400,640){\rule[-0.350pt]{1.156pt}{0.700pt}}
\put(404,639){\rule[-0.350pt]{1.385pt}{0.700pt}}
\put(410,638){\rule[-0.350pt]{1.385pt}{0.700pt}}
\put(416,637){\rule[-0.350pt]{1.385pt}{0.700pt}}
\put(422,636){\rule[-0.350pt]{1.385pt}{0.700pt}}
\put(428,635){\rule[-0.350pt]{1.445pt}{0.700pt}}
\put(434,634){\rule[-0.350pt]{1.445pt}{0.700pt}}
\put(440,633){\rule[-0.350pt]{1.445pt}{0.700pt}}
\put(446,632){\rule[-0.350pt]{1.445pt}{0.700pt}}
\put(452,631){\rule[-0.350pt]{1.385pt}{0.700pt}}
\put(457,630){\rule[-0.350pt]{1.385pt}{0.700pt}}
\put(463,629){\rule[-0.350pt]{1.385pt}{0.700pt}}
\put(469,628){\rule[-0.350pt]{1.385pt}{0.700pt}}
\put(475,627){\rule[-0.350pt]{1.847pt}{0.700pt}}
\put(482,626){\rule[-0.350pt]{1.847pt}{0.700pt}}
\put(490,625){\rule[-0.350pt]{1.847pt}{0.700pt}}
\put(497,624){\rule[-0.350pt]{1.927pt}{0.700pt}}
\put(506,623){\rule[-0.350pt]{1.927pt}{0.700pt}}
\put(514,622){\rule[-0.350pt]{1.927pt}{0.700pt}}
\put(522,621){\rule[-0.350pt]{1.847pt}{0.700pt}}
\put(529,620){\rule[-0.350pt]{1.847pt}{0.700pt}}
\put(537,619){\rule[-0.350pt]{1.847pt}{0.700pt}}
\put(545,618){\rule[-0.350pt]{2.891pt}{0.700pt}}
\put(557,617){\rule[-0.350pt]{2.891pt}{0.700pt}}
\put(569,616){\rule[-0.350pt]{2.770pt}{0.700pt}}
\put(580,615){\rule[-0.350pt]{2.770pt}{0.700pt}}
\put(592,614){\rule[-0.350pt]{2.891pt}{0.700pt}}
\put(604,613){\rule[-0.350pt]{2.891pt}{0.700pt}}
\put(616,612){\rule[-0.350pt]{2.770pt}{0.700pt}}
\put(627,611){\rule[-0.350pt]{2.770pt}{0.700pt}}
\put(639,610){\rule[-0.350pt]{2.770pt}{0.700pt}}
\put(650,609){\rule[-0.350pt]{2.770pt}{0.700pt}}
\put(662,608){\rule[-0.350pt]{5.782pt}{0.700pt}}
\put(686,607){\rule[-0.350pt]{2.770pt}{0.700pt}}
\put(697,606){\rule[-0.350pt]{2.770pt}{0.700pt}}
\put(709,605){\rule[-0.350pt]{5.782pt}{0.700pt}}
\put(733,604){\rule[-0.350pt]{5.541pt}{0.700pt}}
\put(756,603){\rule[-0.350pt]{5.782pt}{0.700pt}}
\put(780,602){\rule[-0.350pt]{5.541pt}{0.700pt}}
\put(803,601){\rule[-0.350pt]{11.322pt}{0.700pt}}
\put(850,600){\rule[-0.350pt]{11.322pt}{0.700pt}}
\put(897,599){\rule[-0.350pt]{56.371pt}{0.700pt}}
\put(1131,600){\rule[-0.350pt]{17.104pt}{0.700pt}}
\put(1202,601){\rule[-0.350pt]{11.081pt}{0.700pt}}
\put(1248,602){\rule[-0.350pt]{11.322pt}{0.700pt}}
\put(1295,603){\rule[-0.350pt]{11.322pt}{0.700pt}}
\put(1342,604){\rule[-0.350pt]{5.782pt}{0.700pt}}
\put(1366,605){\rule[-0.350pt]{11.322pt}{0.700pt}}
\put(1413,606){\rule[-0.350pt]{5.541pt}{0.700pt}}
\end{picture}

\caption[The mass differences as a function of the matching scale~$\Lambda$.
Plotted are $\dem{\pi}$ and the correction to Dashen's theorem.]
{The mass differences as a function of the matching scale~$\Lambda$.
Plotted are $\dem{\pi}$ and the correction to Dashen's theorem, $\Delta D =
\left(m_{K^+}^2 - m_{K^0}^2 -m_{\pi^+}^2 + m_{\pi^0}^2\right)_{em}$.}
\label{fig4}
\end{figure}
\end{document}